\documentclass[aps,twocolumn,nofootinbib,superscriptaddress,prd]{revtex4-2}
\usepackage{graphicx}
\usepackage{amsmath}
\usepackage{amssymb}
\usepackage{amsfonts}
\usepackage{mathtools}
\usepackage{color}
\usepackage{dsfont}
\usepackage{soul}
\usepackage{hyperref}
\usepackage{tikz}
\usetikzlibrary{positioning,shapes}
\DeclareMathOperator{\sinc}{sinc}

\usepackage{changes}
\usepackage{siunitx}
\definechangesauthor[name={Ale}, color=red]{Ale}
\definechangesauthor[name={David}, color=orange]{David}

\usepackage{tikz,xcolor}
\definecolor{lime}{HTML}{A6CE39}
\DeclareRobustCommand{\orcidicon}{%
	\begin{tikzpicture}
	\draw[lime, fill=lime] (0,0) 
	circle [radius=0.16] 
	node[white] {{\fontfamily{qag}\selectfont \tiny ID}};
	\draw[white, fill=white] (-0.0625,0.095) 
	circle [radius=0.007];
	\end{tikzpicture}
	\hspace{-2mm}
}
\foreach \x in {A, ..., Z}{%
	\expandafter\xdef\csname orcid\x\endcsname{\noexpand\href{https://orcid.org/\csname orcidauthor\x\endcsname}{\noexpand\orcidicon}}
}

\begin{document}
\title{
Interplay between optomechanics and the dynamical Casimir effect 
}
\date{\today}

\author{Alessandro Ferreri\orcidA{}}
\affiliation{Institute for Quantum Computing Analytics (PGI-12), Forschungszentrum J\"ulich, 52425 J\"ulich, Germany}
\author{Hannes Pfeifer\orcidE{}}
\affiliation{Institute of Applied Physics, University of Bonn, 53115 Bonn, Germany}
\author{Frank K. Wilhelm}
\affiliation{Institute for Quantum Computing Analytics (PGI-12), Forschungszentrum J\"ulich, 52425 J\"ulich, Germany}
\author{Sebastian Hofferberth\orcidD{}}
\affiliation{Institute of Applied Physics, University of Bonn, 53115 Bonn, Germany}
\author{David Edward Bruschi\orcidB{}}
\affiliation{Institute for Quantum Computing Analytics (PGI-12), Forschungszentrum J\"ulich, 52425 J\"ulich, Germany}

\begin{abstract}
We develop a model of a quantum field confined within a cavity with a movable wall where the position of the wall is quantized. We obtain a full description of the dynamics of both the quantum field and the confining wall depending on the initial state of the whole system. Both the reaction and back-reaction of the field on the wall, and the wall on the field, can be taken into account, as well as external driving forces on both the cavity and the wall.
The model exactly reproduces the resonant cavity mode stimulation due to the periodic motion of the mirror (dynamical Casimir effect), as well as the standard radiation pressure effects on the quantized wall (optomechanics). The model also accounts for the interplay of the two scenarios. Finally, the time evolution of the radiation force shows the interplay between static and dynamical Casimir effect.
\end{abstract}

\maketitle

\section{Introduction}
Optomechanics is the study of the light-matter interaction, which focuses its effort on the study of the interaction between the modes of the electromagnetic field and the vibrational modes of macroscopic bodies, such as reflective mirrors \cite{https://doi.org/10.1002/andp.201200226, doi:10.1126/science.1156032, Kippenberg:07, Aspelmeyer:Kippenberg:2014}. The employment of optomechanical systems for technological purposes \cite{ barzanjeh2021optomechanics}, such as laser cooling \cite{PhysRevA.77.033804, PhysRevLett.99.093902, chan2011laser, teufel2011sideband, PhysRevA.98.023860}
and quantum sensing \cite{PhysRevA.96.043824, RademacherMillenLi+2020+227+239}, as well as the possibility to involve macroscopic objects in the investigation of quantum physics, such as studies of entanglement \cite{doi:10.1126/science.1211914, riedinger2018remote}, has lead to a growing interest in this field throughout the last two decades \cite{doi:10.1063/1.5106409,doi:10.1063/1.4896029, Vanner16182}.

The optomechanical coupling arises due to a nonlinear interaction wherein the excitation (or relaxation) of the vibrational mode of the mirror is determined by the radiation pressure induced by the presence of photons within the cavity. 
The most common approach to the analysis of optomechanical interactions treats the mirror and the electromagnetic field as two independent quantum harmonic oscillators \cite{Qvarfort_2019,Qvarfort_2020, PhysRevResearch.3.013159}. 
In this approach, modelling the electromagnetic field as a single harmonic oscillator turns out to be extremely helpful for the description of the effects of cavity losses \cite{PhysRevA.104.013501}. Moreover, it can be employed in order to investigate phenomena of vacuum squeezing acting on either the mechanical mode \cite{Bruschi_2018, PhysRevA.100.063827} or the cavity field. The latter case is known as the dynamical Casimir effect (DCE) \cite{physics2010007}, which can be understood through the lens of optomechanics as the resonant exchange of quantum excitations between the mechanical mode and one, or two, cavity modes of the electromagnetic field \cite{ PhysRevX.8.011031, PhysRevA.100.062501, PhysRevA.100.022501, PhysRevA.100.062516}.

The single mode description of the field trapped in a cavity conveniently simplifies the study of the system in many scenarios. Nevertheless, the investigation of the multimode nature of the quantum field in the framework of optomechanics leads to a more comprehensive interpretation of the interplay between the quantum degrees of freedom of the wall and the quantum fluctuation of the field \cite{PhysRevA.51.2537}. Concretely, this translates into the quantization of the position of the wall that is classically determined by Dirichlet boundary conditions for the field \cite{Alves_2003, Dalvit_2006, PhysRevA.64.013808}, and subsequently the quantization of the fluctuations of the length which become coupled to the creation and annihilation operators of the cavity modes \cite{PhysRevA.51.2537}. 
This additional quantum degree of freedom can be exploited to investigate many interesting features of this system, such as the correction to both the vacuum state and its energy of the multimode cavity field caused by the motion of the wall \cite{PhysRevLett.111.060403, PhysRevD.91.025012, PhysRevD.96.045007}. 

In this work we propose an approach to extend the standard optomechanical radiation-pressure coupling to include the mode-mixing and squeezing field terms in the Hamiltonian of the system. We specialize our computations to a real scalar field fulfilling static Dirichlet boundary conditions without loss of generality. Our goal is achieved by adding a small fluctuation to the length of the cavity, corresponding to the oscillation amplitude of the wall, and then performing the spatial integration of the Hamiltonian density to lowest order in the fluctuation length. This additional degree of freedom is then quantized, and the resulting quantum Hamiltonian reproduces the standard optomechanical one with the addition of all cavity modes \cite{PhysRevA.51.2537}.

We compute the time evolution of the whole system including possible time-dependent external drives acting on two different modes of the cavity field, as well as on the mechanical degree of freedom. Moreover, we consider an input state that accounts for the effects of displacement, squeezing and thermal fluctuation of the mechanical mode \cite{PhysRevA.47.4474}. One of our most interesting results is the excitation of the vacuum state of the cavity mode by means of a time-dependent displacement acting on the mechanical degree of freedom. We demonstrate that our model allows to formally recover the DCE as a result of a time-dependent external drive, without necessarily introducing the dynamics into the boundary conditions of the field. Furthermore, our approach allows us not only to recover well-known results, such as the DCE caused by both the resonant oscillating wall and thermal fluctuations of the mechanical state \cite{PhysRevA.100.062516}, but also to unveil novel features, such as the DCE in the presence of squeezed thermal input states for the mechanical mode, the impact of external drives on the dynamics, as well as to estimate the Casimir force between the two cavity walls throughout the dynamics.

The paper is structured as follows. In Section~\ref{theoretical:model}  we introduce the system of interest. In particular, starting from the Lagrangian of the confined field, we present the quantization process, as well as the time evolution formalism. In Section \ref{Dynamics}, we describe the dynamics of relevant quantities, such as the position of the moving wall, the number of photons within the cavity and the force between the two mirrors composing the cavity, showing the interplay between the Casimir force and the time-dependent radiation pressure due to the arising number of photons. In Section \ref{appext} we discuss a list of possible experimental realisations of our system as well as an extension of our model to a scalar massive field. In Section \ref{consout} we include some considerations as well as an outlook. Finally, we summarize in Section \ref{conclusion}.

\section{Theoretical model}\label{theoretical:model}
Here we introduce the model that will be used in this work. Standard approaches to quantum fields confined in cavities can be found in the literature \cite{PhysRevA.51.2537}. 

\subsection{Classical Lagrangian and Hamiltonian density}
Let us consider a massless real scalar field $\phi(t,x)$ in $1+1$ dimensional flat spacetime with coordinates $(t,x)$.\footnote{We use Einstein's summation convention.} The classical Lagrangian for the field density is
\begin{align}\label{lagrangian:density}
\mathcal{L}(t,x)=\frac{\hbar}{2}\partial_\mu\phi\partial^\mu\phi.
\end{align}
The Lagrangian provides the classical field equation, i.e., the Klein Gordon equation, which reads $\partial_t^2\phi-\partial_x^2\phi$=0. We then impose the static Dirichlet boundary conditions $\phi(t,0)=\phi(t,L)=0$ to implement the confinement of the field in a cavity of length $L$.

Since we are interested in the time evolution of the system, we introduce a basis for the field modes $\phi_n(t,x)=\sqrt{\frac{c}{ \omega_n L}}e^{-i\omega_n t}\sin(\frac{n\pi}{L}x)$, where $\omega_n:=\frac{n\pi c}{L}$ is the frequency of mode $n\in\mathbb{N}$. The field can therefore be expanded as
\begin{align}\label{field:expression}
\phi(t,x)=&\sum_{n=1}\left[\alpha_n\,\phi_n(t,x)+\alpha_n^*\,\phi^*_n(t,x)\right],
\end{align}
where the constants $\alpha_n$ are Fourier coefficients.

We can construct the Hamiltonian density $\mathcal{H}:=\Pi\partial_t\phi-\mathcal{L}$ by firstly finding the conjugate momentum $\Pi(t,x):=-\partial_t\phi(t,x)$, then by substituting the expression of $\partial_t\phi$ in terms of the momentum into $\mathcal{H}$, thereby obtaining $\mathcal{H}(t,x)=\frac{1}{2}\left[\Pi^2(t,x)+(\partial_x\phi(t,x))^2\right]$. Finally, in the Schrödinger picture the explicit expression in term of the field expansion coefficients therefore reads
\begin{align}\label{hamiltonian:density:explicit}
\mathcal{H}=&-\frac{\hbar}{2L}\sum_{nm}\sqrt{\omega_n\omega_m}\left[\alpha_n-\alpha_n^*\right]\left[\alpha_m-\alpha_m^*\right]s_n(x)s_m(x)\nonumber\\
&+\frac{\hbar}{2L}\sum_{nm}\sqrt{\omega_n\omega_m}\left[\alpha_n+\alpha_n^*\right]\left[\alpha_m+\alpha_m^*\right]c_n(x)c_m(x)\nonumber\\
\end{align}
where we have introduced $s_n(x):=\sin(\frac{n\pi}{L}x)$ and $c_n(x):=\cos(\frac{n\pi}{L}x)$ here only for the sake of presentation.

\subsection{Quantization}\label{quantization:H}
\begin{figure}[h]
	\centering
	\includegraphics[width=0.9\linewidth]{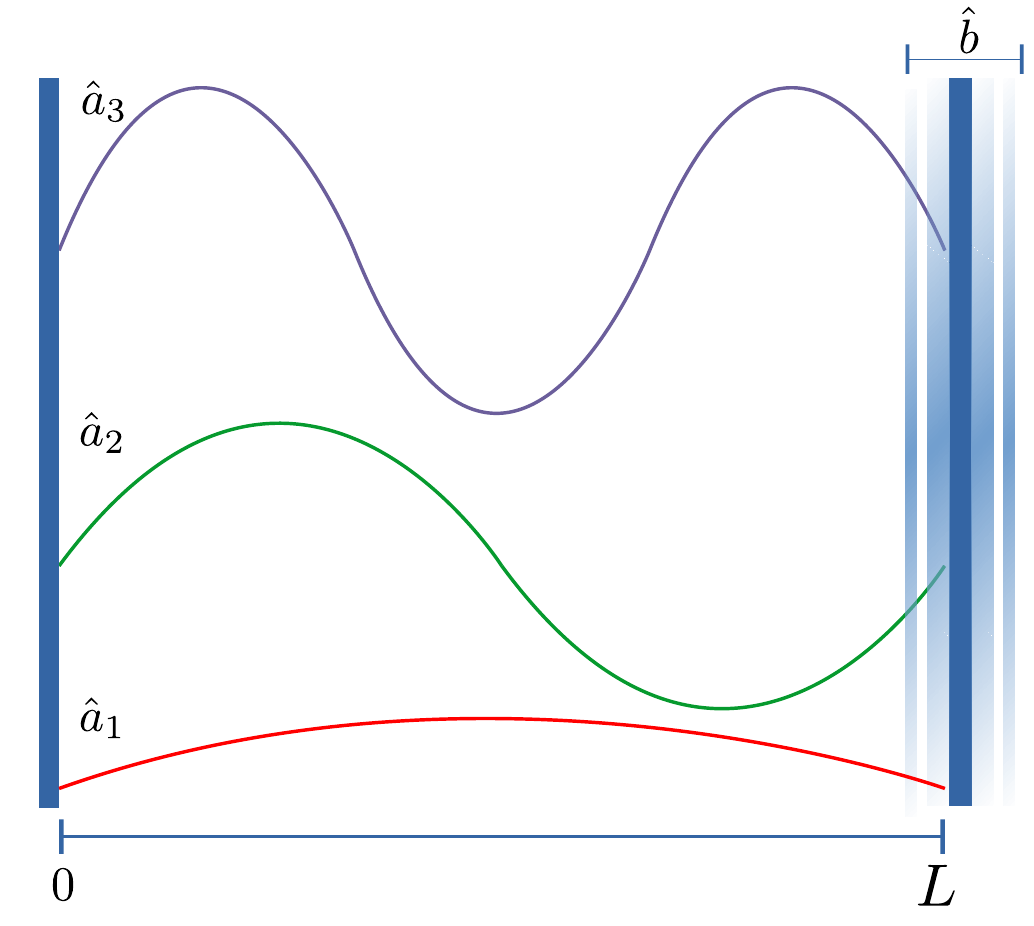}
	\caption{\textbf{Schematic picture of the system}. A 1+1 dimensional scalar field is confined in a cavity of length $L$ which possesses a movable wall. The fundamental oscillation amplitude is proportional to the zero point fluctuation $\delta L_0$ of the quantum harmonic oscillator.}
	\label{system}
\end{figure}

So far we have dealt with a classical system. We now wish to quantize the system, and in particular the Hamiltonian, with classical Hamiltonian density \eqref{hamiltonian:density:explicit}. To obtain our goal we adopt the following four-step strategy:
\begin{itemize}
    \item[i)] In Equation~\eqref{hamiltonian:density:explicit} we substitute $L\rightarrow L+\delta L$, where $\delta L/L\ll1$, and Taylor-expand up to the first order in $\delta L/L$;
    \item[ii)] We integrate the Hamiltonian density from 0 to $L$ and obtain the classical Hamiltonian $H:=\int_0^L \mathcal{H}\, dx$;
    \item[iii)] We quantize both the field Fourier coefficients $\alpha_n$ and the position of the wall $\delta L$:
\begin{align}\label{hamiltonian:explicit}
\alpha_n\rightarrow&\hat{a}_n\nonumber,\\
\alpha_n^*\rightarrow&\hat{a}^\dag_n\nonumber,\\
\delta L\rightarrow& \delta L_0 (\hat{b}^\dag+\hat{b}). 
\end{align}
Here $\delta L_0$ is a constant and represents the zero-point fluctuations of the harmonic oscillator $\hat{b}$ (details on this topic are left to the literature \cite{ Aspelmeyer:Kippenberg:2014}), and the operators all satisfy the canonical commutation relations $[\hat a_n,\hat a_{n'}^\dag]=\delta_{nn'}$ and $[\hat{b},\hat{b}^\dag]=1$, while all other commutators vanish. We have therefore assumed that the (small) displacement of one of the two cavity walls is characterized by a harmonic degree of freedom with intrinsic frequency $\omega$ and annihilation and creation operators $\hat b$ and $\hat b^\dagger$ respectively.
\item[iv)] Finally, we normal order the quantum Hamiltonian $\hat{H}$ that we obtained this way and we introduce the dimensionless amplitude $\epsilon:=\delta L_0/L\ll1$ which will act as our perturbative parameter.
\end{itemize}
The system is pictorially illustrated in Figure~\ref{system}.

The procedure described here  allows us to obtain the total Hamiltonian $\hat{H}(t)$ of the form
\begin{align}\label{quantum:hamiltonian:explicit}
\hat{H}(t)=&\hat H_0+\hat H_{\textrm{dr}}(t)+\epsilon\hat H_{\textrm{I}},
\end{align}
where each term reads
\begin{align}\label{quantum:hamiltonian:terms2}
\hat{H}_0:=&\sum_{n}\hbar\omega_n\,\hat{a}_n^\dag\hat{a}_n+\hbar\omega\hat{b}^\dag\hat{b}\nonumber,\\
\hat{H}_{\textrm{dr}}(t):=&\hat H_{\textrm{dk}}(t)+\hat H_{\textrm{dk'}}(t)+\hat H_{\textrm{db}}(t),\nonumber\\
\hat{H}_{\textrm{I}}:=&-2\sum_{n}\hbar\omega_n\,\hat{a}_n^\dag\hat{a}_n \hat X_{b}-\sum_{n}\hbar\omega_n\,\left(\hat{a}_n^{\dag2}+\hat{a}_n^2\right)\hat X_{b}\nonumber\\
&-4\sum_{\substack{n,m\\n\neq m}}(-1)^{n+m}\hbar\sqrt{\omega_n\omega_m}\,\hat X_n\hat X_{m}\hat X_{b}.
\end{align}
Here, the added term $\hat H_{\textrm{dr}}(t)$
includes potential external driving forces that can be applied to both the cavity field, for example to the modes $k$ and $k'$, and the mechanical mode $b$, 
$\hat H_{\textrm{dj}}(t)=2\lambda_{xj}(t)\hat X_j+2\lambda_{pj}(t)\hat P_j$ with $j=k, k'$ or $b$.
We have also compacted our notation by introducing the quadrature position and momentum operators $\hat X_{b}=\frac{1}{2}(\hat b^\dagger+\hat b)$, $\hat P_{b}=\frac{i}{2}(\hat b^\dagger-\hat b)$, $\hat X_{n}=\frac{1}{2}(\hat a_n^\dagger+\hat a_n)$, and $\hat P_{n}=\frac{i}{2}(\hat a_n^\dagger-\hat a_n)$.

We want to emphasize that we perform both the substitution and the Taylor expansion with respect to $\delta L/L$ \emph{before} the spatial integration of the Hamiltonian density. Interestingly, this approach spontaneously provides both the standard optomechanical interaction Hamiltonian  $\hat{H}_{\textrm{OM}}=-2\epsilon\sum_{n}\omega_n\,\hat{a}_n^\dag\hat{a}_n\hat X_b$ commonly found in the literature \cite{Aspelmeyer:Kippenberg:2014}, and the cavity multimode mono- and two-colour squeezing interaction Hamiltonian $H_{\textrm{sq}}=-\sum_{n}\hbar\omega_n\,\left(\hat{a}_n^{\dag2}+\hat{a}_n^2\right)\hat X_{b}-4\sum_{n\neq m}(-1)^{n+m}\hbar\sqrt{\omega_n\omega_m}\,\hat X_n\hat X_{m}\hat X_{b}$, which induces the dynamical Casimir effect \cite{PhysRevA.51.2537, PhysRevLett.111.060403}.

\subsection{Time evolution operator}
In the last section we derived the Hamiltonian of the system. In general, such Hamiltonian can depend on time either due to the modulation of the coupling constants $g_n:=\epsilon\omega_n$ (which has been considered in the literature \cite{Aspelmeyer:Kippenberg:2014, PhysRevLett.119.053601}), or the presence of time-dependent external drives \cite{PhysRevLett.122.030402}. Here we do not consider coupling modulation and instead investigate the dynamics of a system described by the Hamiltonian \eqref{quantum:hamiltonian:explicit}.
The time evolution operator $\hat{U}(t)$ induced by a time-dependent Hamiltonian $\hat{H}(t)$ reads
\begin{align}\label{time:evolution:operator}
\hat{U}(t)=\overset{\leftarrow}{\mathcal{T}}\exp\left[-\frac{i}{\hbar}\int_0^{t}\,dt'\,\hat{H}(t')\right],
\end{align}
where $\overset{\leftarrow}{\mathcal{T}}$ stands for the time-ordering operator.

It is not difficult to show that
\begin{align}\label{time:evolution:operator}
\hat{U}(t)=\hat{U}_0(t)\hat{U}_{\textrm{dr}}^{(+)}(t)\hat{U}_{\textrm{dr}}^{(-)}(t)\hat{U}_{\textrm{I}},
\end{align}
modulo an overall complex phase that has no physical significance. This expression is specified by the following unitary operators 
\begin{align}\label{quantum:hamiltonian:terms}
\hat{U}_0(t):=&\exp\left[-i/\hbar\,\hat{H}_0 t\right],\\ 
\hat{U}_{\textrm{dr}}^{(+)}(t):=&\exp\left\{-2i\left[\Lambda_{xb}(t)\hat X_b+\Lambda_{xk}(t)\hat X_k\right.\right.\nonumber\\
&\left.\left.+\Lambda_{xk'}(t)\hat{X}_{k'}\right]\right\},
\\ 
\hat{U}_{\textrm{dr}}^{(-)}(t):=&\exp\left\{-2i\left[\Lambda_{pb}(t)\hat P_b+\Lambda_{pk}(t)\hat P_k\right.\right.\nonumber\\
&\left.\left.+\Lambda_{pk'}(t)\hat P_{k'}\right]\right\},
\\ 
\hat{U}_{\textrm{I}}(t):=&\overset{\leftarrow}{\mathcal{T}}e^{-\frac{i}{\hbar}\epsilon\int_0^t dt'\hat{\tilde{H}}_{\textrm{I}}(t')}.
\end{align}
We also need the definition of the auxiliary functions
{\small\begin{align}
\Lambda_{pj}(t):=&\int_0^tdt'\left(\lambda_{xj}(t')\sin(\omega t')-\lambda_{pj}(t')\cos(\omega t')\right),\nonumber\\
\Lambda_{xj}(t):=&\int_0^tdt'\left(\lambda_{xj}(t')\cos(\omega t')+\lambda_{pj}(t')\sin(\omega t')\right),
\end{align}
}
and $j\equiv k,k'$ or $b$, and we also have
\begin{equation}
\hat{\tilde{H}}_{\textrm{I}}(t):=\hat{U}^\dag_{\textrm{d}}(t)\hat{U}^\dag_0(t)\hat{H}_{\textrm{I}}(t)\hat{U}_0(t)\hat{U}_{\textrm{d}}(t),
\label{H:tilde}
\end{equation}
with $\hat{U}_{\textrm{d}}(t):=\hat{U}_{\textrm{dr}}^{(+)}(t)\hat{U}_{\textrm{dr}}^{(-)}(t)$.
The explicit expression of $\hat{\tilde{H}}_{\textrm{I}}(t)$ is cumbersome and is provided in Appendix~\ref{app:tools}.

\subsection{Time evolution and the initial state}
We are now able to determine the time evolution of the expectation value of any observable $\hat{A}$. In the Heisenberg picture, this is given by 
\begin{align}
A(t)=\textrm{Tr}(\hat{U}^\dag(t)\hat{A}(0)\hat{U}(t)\hat{\rho}(0)),
\label{time:evolution:o}
\end{align}
where $\hat{\rho}(0)$ is the initial state of the total cavity field and wall system. We choose to work with the following general initial state:
\begin{align}
\hat{\rho}(0)=\lvert\mu_{k}\rangle\langle\mu_{k}\rvert\otimes\lvert\mu_{k'}\rangle\langle\mu_{k'}\rvert\otimes\hat{\rho}^{\textrm{rest}} \otimes\hat{\rho}_{\textrm{m}}^{\textrm{(DST)}},
\label{initial:state:complete}
\end{align}
where $\hat{\rho}^{\textrm{rest}}:=\prod_{n\neq k,k'}\lvert 0_n\rangle\langle 0_n\rvert$ is the vacuum state of all modes except two modes $k,k'$. Concretely, we have assumed that two specific cavity modes $k,k'$ are initially found in a coherent state defined by $\hat{a}_k|\mu_k\rangle=\mu_k|\mu_k\rangle$ and with coherent parameter $\mu_k\in\mathbb{R}$ (analogously for $k'$), whereas the mechanical quantum mode of the wall is prepared in a displaced squeezed thermal (DST) state $\hat{\rho}_{\textrm{m}}^{\textrm{(DST)}}=D(\beta)S(\zeta)\rho_m(T)S^\dagger(\zeta)D^\dag(\beta)$, where T is the temperature, and $\beta=\lvert\beta\rvert e^{i\theta}$ and $\zeta=re^{i\phi}$ are the coherent and the squeezing parameter of the mechanical degree of freedom respectively \cite{PhysRevA.47.4474}. 

The total number of excitations at $t=0$ is therefore $N_{tot}(0)=N_{k}(0)+N_{k'}(0)+N_b(0)$,
where $N_{k}(0)=\mu_k^2$, $N_{k'}(0)=\mu_{k'}^2$, and
\begin{align}
N_b(0)=\lvert\beta\rvert^2+\sinh^2r+N_T\cosh(2r),
\label{nb0}
\end{align}
with $N_T=\sinh^2(r_{\textrm{T}})$ and the parameter $r_{\textrm{T}}$ defined by $\tanh r_{\textrm{T}}:=\exp[-\frac{\hbar\omega}{2 k_{\textrm{B}} T}]$.

Concretely, we can compute the time evolution of $\hat{A}$ by Taylor expanding the unitary operator $U_I(t)$ up to the second order in $\epsilon$,
{\small
\begin{align}
U_I(t)\simeq& 1-\frac{i\epsilon}{\hbar}\int_0^t dt'\hat{\tilde{H}}_{\textrm{I}}(t')-\frac{\epsilon^2}{\hbar^2}\int_0^t dt'\hat{\tilde{H}}_{\textrm{I}}(t')\int_0^{t'} dt''\hat{\tilde{H}}_{\textrm{I}}(t'').
\label{unit.op.:expansion}
\end{align}
}
From now on, we assume that all quantities are truncated at second order.

Following this procedure we obtain
\begin{align}
A(t)\simeq& A^{(0)}(t)+\frac{i}{\hbar}\epsilon \Delta A^{(1)}(t)+\frac{1}{\hbar^2}\epsilon^2 \Delta A^{(2)}(t),
\end{align}
where each term has the expression
{\small\begin{align}
A^{(0)}(t)=& \textrm{Tr}(\hat{\tilde{A}}(t)\hat{\rho}(0)),\nonumber\\
A^{(1)}(t)=&\textrm{Tr}\left(\left[\int_0^tdt'\hat{\tilde{H}}_{\textrm{I}}(t'),\hat{\tilde{A}}(t)\right]\hat{\rho}(0)\right),\nonumber\\
A^{(2)}(t)=&-\textrm{Tr}\left(\int_0^t dt'\int_0^{t'} dt''\hat{\tilde{H}}_{\textrm{I}}(t'')\hat{\tilde{H}}_{\textrm{I}}(t') \hat{\tilde{A}}(t)\hat{\rho}(0)\right)\nonumber\\
&-\textrm{Tr}\left(\int_0^t dt'\int_0^{t'} dt''\hat{\tilde{A}}(t)\hat{\tilde{H}}_{\textrm{I}}(t')\hat{\tilde{H}}_{\textrm{I}}(t'')\hat{\rho}(0)\right)\nonumber\\
&+\textrm{Tr}\left(\int_0^tdt'\hat{\tilde{H}}_{\textrm{I}}(t')\hat{\tilde{A}}(t)\int_0^tdt''\hat{\tilde{H}}_{\textrm{I}}(t'')\hat{\rho}(0)\right),
\label{time:evolution:second:order}
\end{align}
}
and where $\hat{\tilde{A}}(t):=\hat{U}_{\textrm{d}}^\dag(t)\hat{U}_0^\dag(t)\hat{A}(t)\hat{U}_0(t)\hat{U}_{\textrm{d}}(t)$.

\subsection{Time average of measurable quantities}
We anticipate that the observables computed in our work will depend on time through complicated oscillatory functions. While this functional dependence on time can apparently obfuscate the physical significance, we can extract meaningful significance by taking the time average of such quantities. This is particularly useful because the time average in general eliminates oscillations around mean values.

Given the expectation vale $A(t)$ of an observable $\hat{A}(t)$, we define the \textit{time average} $\langle A\rangle_\tau$ of the expectation value $A(t)$ as 
\begin{align}
\langle A\rangle_\tau:=\frac{1}{\tau}\int_0^\tau dt\,A(t).
\end{align}
These time averaged quantities will be computed case by case below.

\section{Dynamics of the system}\label{Dynamics}
In the following analysis, we focus our investigation on three quantities: the position of the oscillating wall, the number of photon for the single mode k, and the radiation pressure within the cavity.

\subsection{Dynamics of the wall}
We start by studying the dynamics of the wall. We are therefore interested in calculating the average position $x(t)$ of the wall at time t. This is given by
\begin{align}
x(t)=L+\delta L_0\textrm{Tr}(\hat{U}^\dag(t)\hat (\hat b+\hat b^\dag)\hat{U}(t)\hat{\rho}(0)),
\end{align}
which has the general expression $x(t)=x^{(0)}(1+\epsilon \tilde{x}^{(1)}+\epsilon^2 \tilde{x}^{(2)})$. Notice here that, since the length has a dimension, it must be rescaled in order to have dimensionless perturbative contributions. In the following we consider the terms $\tilde{x}^{(n)}$ as the rescaled corrections to the zero order.
In our frame of reference, the second wall, which remains static, is assumed to be located at $x=0$.

\subsubsection{Zero order: $x^{(0)}(t)$}\label{sect:x:zero}
We are now able to investigate the dynamics of the wall starting from the zero order contribution.
It is immediate to show that the trajectory of the movable mirror at lowest order is
\begin{align}
x^{(0)}(t)=&L.
\label{x:zero:order}
\end{align}
This is expected and, for our purposes, not of particular significance.

\subsubsection{First order: $\tilde{x}^{(1)}(t)$}\label{sect:x:first}
We continue by looking at the first order, i.e., the terms that contribute as a consequence of the back reaction of the field on the cavity wall. It is easy to find
\begin{align}
\tilde{x}^{(1)}(t)=&2\lvert\beta\rvert\cos(\omega t-\theta)+2\xi(t),
\label{x:first:order}
\end{align}
where the function $\xi(t)$ is defined in Appendix \ref{app:tools}.

The dynamics is triggered by two terms. In particular, the first one accounts for the harmonic oscillation of the mirror, and directly stems from the quantization of its mechanical motion; indeed, the oscillation amplitude is strictly connected to the number of phonons due to the factor $\lvert\beta\rvert$, whereas both the initial position and velocity are established by the coherent phase $\theta$, which also determines the orientation of the phonon momentum. There are a few interesting values that the phase can assume. For instance, if $\theta=0$ (or $\pi$), the mirror is initially pushed (or pulled) at maximal amplitude $\delta L_0\lvert\beta\rvert$ and the canonical momentum of phonons is zero. If $\theta=\pm\pi/2$ instead, the mirror initially moves at maximal velocity, and the phonons have positive ($\theta=+\pi/2$) or negative ($\theta=-\pi/2$) momentum.
The second term is the contribution of the mechanical drive to the motion of the wall, and its explicit expression depends on our choice of $\lambda_{xb}(t)$ and $\lambda_{pb}(t)$. 

Let us define the external drive as follows:
\begin{align}
\lambda_{xb}(t)=&-\frac{g\Omega}{2}e^{-\Omega t}\cos(\omega t),\nonumber\\
\lambda_{pb}(t)=&-\frac{g\Omega}{2}e^{-\Omega t}\sin(\omega t),
\label{exdr}
\end{align}
where $g$ is the coupling constant of the external drive and $\Omega$ has units of frequency.
Accordingly, the equation of motion \eqref{x:first:order} becomes
\begin{align}
\tilde{x}^{(1)}(t)= &2\lvert\beta\rvert\cos(\omega t-\theta)+g\sin\left(\omega t\right)\left(1-e^{-\Omega t}\right).
\label{x:oscill}
\end{align}
The specific choice of the external drive generates a sinusoidal modulation of the wall when $\Omega t\gg 1$; moreover, it ensures both $\tilde{x}^{(1)}(0)=0$ and $\dot{\tilde{x}}^{(1)}(0)=0$ when $\beta=0$, namely a smooth transition from the static to the dynamical regime. The constant $g$ determines both the intensity and the initial direction of the modulation. Throughout this paper we will always assume two conditions: $\Omega\neq\omega_n$ and $\Omega\gg \omega$. The first condition prevents any resonances between the frequency $\Omega$ and any cavity modes of the field, whereas the second one allows us to neglect the exponential decay $e^{-\Omega t}$ in our formulas.

\subsubsection{Second order: $\tilde{x}^{(2)}(t)$}\label{section:second:x}
We proceed to the second order corrections. Unfortunately, the presence of cavity drives makes the second order corrections quite difficult to compute, therefore we opt for discussing the scenario without cavity drives for this case. We report the full formula \eqref{xcompl} in Appendix~\ref{x:second:order}.
Thus, henceforth we impose $\lambda_{xk}(t)=\lambda_{xk'}(t)=\lambda_{pk}(t)=\lambda_{pk'}(t)\equiv 0$.

We will focus our analysis on the specific case when one of the possible resonances between the cavity modes $\omega_n$ and the mechanical frequency $\omega$ occurs. From Equation~\eqref{xcompl} we identify three possible resonance conditions:
\begin{itemize}
\item[i)] Degenerate resonance (also known as single mode squeezing): $\omega= 2\omega_k$;
\item[ii)] Nondegenerate resonance (also known as two mode squeezing): $\omega= \omega_k+\omega_{k'}$;
\item[iii)] Intermode coupling (also known as mode mixing): $\omega=\omega_k-\omega_{k'}$.
\end{itemize}
We stress that only a multimode description of the cavity field as the one presented here can take the second and the third resonances into account \cite{physics2010007, PhysRevA.64.013808}.

We assume to operate in the degenerate case i), where $\omega=2\omega_k$ for a chose mode $k$. This choice allows us to employ  Equation~\eqref{xcompl} to obtain
\begin{align}
\tilde{x}^{(2)}(t)\simeq &\mu_{k}^2\left(1-\cos\left(2\omega_k t\right)+\omega_k t\sin(2\omega_k t)\right).
\label{x:first:resonance}
\end{align}
In this formula all off-resonant terms were neglected, since they are by a factor $\omega_k$ or even $\omega_k^2$ smaller than the resonant ones.

We note that also the second order contributions \eqref{x:first:resonance} show a linear growth of the oscillation amplitude. The occurrence of this linear growth stems specifically from both the choice of an initial coherent state for the cavity mode and the resonance between the excited cavity mode and mechanical oscillation.

The time average of the correction $\tilde{x}^{(2)}(t)$ for large enough times reads
\begin{align}
\langle\tilde{x}^{(2)}\rangle_\tau= &\mu_{k}^2\left(1-\frac{1}{2}\cos(2\omega_k \tau)\right).
\end{align}
Interestingly, we note that since the $|\cos \alpha|\leq1$, we have that $\langle\tilde{x}^{(2)}\rangle_\tau\geq0$ for all times. This implies that, on average to this order, the wall of the cavity is always pushed outwards, and we trace this effect to the back reaction of resonance between the cavity mode and the harmonic oscillator. This effect is subdominant to the one obtained at first order, which is expected since back reaction on the wall in the form of radiation pressure occurs as a higher order than the direct effect of driving the wall with an external force, or with an initial displacement.

\subsubsection{Role of the phase in the initial wall displacement}\label{section:second:x2}
We are now able to discuss the role played by the phase $\theta$ throughout the dynamics of the system. As already pointed out in Section~\ref{sect:x:zero}, the coherent phase $\theta$ determines both the expected position and momentum of coherent phonons. We now focus on two specific values. 

\textit{Purely real mechanical coherent parameter.} This situation occurs if $\theta=n \pi$ with $n\in\mathbb{Z}$. Manipulating the overall expression for $x(t)$ we find
\begin{align}
x(t)\simeq &L\left[1+ \epsilon\left(2(-1)^n\lvert\beta\rvert-\epsilon\mu_{k}^2\right)\cos(2\omega_k t)+\epsilon^2\mu_{k}^2\right.\nonumber\\
&\left.+\epsilon\sin(2\omega_k t)\left(\epsilon\mu_{k}^2\omega_k t+g\right)\right].
\label{x:first:resonance2}
\end{align}
Although this expression clearly highlights the impossibility to annul the linear growth of the oscillation amplitude, the mechanical drive could partially inhibit it as long as $g$ is negative. Moreover, by properly tuning the coherent parameter $\beta$, we could suppress
the additional beat caused by the term proportional to $\cos(2\omega_k t)$.

\textit{Purely imaginary mechanical coherent parameter.}
We now prepare the harmonic oscillator in a coherent state with purely imaginary parameter: this occurs if $\theta=(n+1/2)\pi$, with $n\in\mathbb{Z}$. This choice allows us to rewrite the equation of motion of the wall as following:
\begin{align}
x(t)\simeq &L\left[1+2\epsilon^2\mu_{k}^2\sin^2\left(\omega_k t\right)\right.\nonumber\\
&+\left.\epsilon\sin(2\omega_k t)\left(\epsilon\mu_{k}^2\omega_k t+2(-1)^n \lvert\beta\rvert+g\right)\right].
\label{x:first:resonance3}
\end{align}
The action of the external drive does not differ from the case reported above, and therefore we can momentarily switch it off.

In order to appreciate the role played by phonons throughout the dynamics in more detail, we distinguish two possible scenarios: either $\theta=+\pi/2$, achieved by assuming $n=0$, merely leading to an increment of the oscillation amplitude; or $\theta=-\pi/2$, namely $n=1$, which leads to a reduction of the oscillation.
We interpret the latter case as follows; phonons, having maximal negative momentum, initially boost the oscillation of the wall leftwards; on the other hand, the resonance between cavity and mechanical mode induces a positive (rightward) push of the wall by means of the radiation pressure within the cavity due to the coherent photons; the damping of the oscillation of the wall in fact ensues from the combination of these two effects, and the damping time is found to be $\bar t^{\,(1)}_f=N_b(0)/\lvert\beta\rvert\mu_k^2\omega_k$ with $\bar t=\epsilon t$. In section \eqref{N:first:order} we will see that the attenuation of the motion is connected with both a reduction in time of the number of phonons, and the corresponding enhance of the photon number.

\subsubsection{Third order: $x^{(3)}(t)$}
In the last section we studied the motion of the wall when one cavity mode is initially found in a coherent state. 
When the cavity field is initially in its vacuum state, i.e., $\mu_k=0$, the first and second order corrections due to the cavity field initial state (namely proportional to the photon number $\mu_k^2$) vanish, and third order effects become dominant. We note that these effects do not arise due to genuine third order contributions, but due to a combination of first and second order dynamical ones. For this reason, it is relevant to compute and analyze them. While it is true that they are greatly suppressed in the perturbative series, they will give us additional information regarding the dynamics.

The initial state for this case is $\hat{\rho}(0)=\prod_{n}\lvert 0_n\rangle\langle 0_n\rvert\otimes\hat{\rho}_{\textrm{m}}^{\textrm{(DST)}}$, where now all the field modes are initially in the vacuum state. 
In this scenario, we present only the case of degenerate resonance where $\omega=2\omega_k$ and no external drive act on the cavity field, $\lambda_{xk}(t)=\lambda_{xk'}(t)=\lambda_{pk}(t)=\lambda_{pk'}(t)\equiv 0$. The correction to the third order therefore reads
{\small\begin{align}
x^{(3)}(t)\simeq&-\frac{\omega_k^2 t^2}{4}\left(2\lvert\beta\rvert \cos(2\omega_k t-\theta)+g\sin(2\omega_k t)\right).
\end{align}
}
Employing this formula within the full expression of the average position gives us the final expression
\begin{align}
\frac{x(t)}{L}\simeq &1+2\epsilon\lvert\beta\rvert\cos\theta\cos(2\omega_k t)\Gamma_k(t)\nonumber\\
+&\epsilon (g+2\lvert\beta\rvert\sin\theta)\sin\left(2\omega_k t\right)\Gamma_k(t),
\label{x:second:resonance}
\end{align}
where we have introduce the important quantity
\begin{align}
    \Gamma_k(t):=&1-\frac{\epsilon^2\omega_k^2 t^2}{4}\approx e^{-\frac{\epsilon^2\omega_k^2 t^2}{4}}.
    \label{decay}
\end{align}
Recall that, since we are working in perturbation theory, our expressions are correct and can be employed for all times where $1-\Gamma_k(t)\ll1$. 

Interestingly, we can immediately observe the gradual reduction of the modulation amplitude since $\Gamma_k(t)$ decreases with time. We will see in section \ref{N:second:order} that this effect stems from the resonant conversion of mechanical excitations (phonons) into photons.
As done before, we want to distinguish two cases of interest, namely when $\beta$ is real or imaginary.

\textit{Real mechanical coherent parameter.}
This occurs when $\theta=n \pi$ with $n\in\mathbb{Z}$. We can rewrite Equation~\eqref{x:second:resonance} as
{\small\begin{align}
\frac{x(t)}{L}\simeq &1+\epsilon\left[2(-1)^n\lvert\beta\rvert\cos(2\omega_k t)+ g\sin\left(2\omega_k t\right)\right]\Gamma_k(t).
\label{x:second:resonance2}
\end{align}
}
We note that the phononic contribution and the action of the drive give rise to two oscillating terms which are dephased by $\pi/2$, and whose amplitude decrease in time.

\textit{Purely imaginary mechanical coherent parameter.} If we assume that $\beta$ is purely imaginary in Equation~\eqref{x:second:resonance}, $\theta=\left(n+\frac{1}{2}\right)$ with $n\in\mathbb{Z}$, we can immediately observe that all contributions oscillate with the same phase:
\begin{align}
\frac{x(t)}{L}\simeq &1+\epsilon\left(g+2(-1)^n\lvert\beta\rvert\right)\Gamma_k(t)\sin\left(2\omega_k t\right).
\label{x:second:resonance3}
\end{align}
This means that we can fully control the oscillation amplitude of the cavity wall. In particular, by properly tuning both the drive coefficient and the phononic coherent parameter, we can amplify or deamplify the oscillation, or even suppress it by imposing $g=-2(-1)^n\lvert\beta\rvert$.
\subsection{Number of photons}
The dynamics of the mirror is strictly connected with the amount of excitations present in the system.
Therefore, we proceed to study the average photon number $N_k(t):= \textrm{Tr}\bigl[\hat U^\dag(t)\hat a_k^\dag\hat a_k \hat U(t)\hat{\rho}(0)\bigr]$ of mode k at time t.
We will make use of the initial state \eqref{initial:state:complete}.

\subsubsection{Zero order: $N_k^{(0)}(t)$}
We start from the lowest order contribution, which provides an accurate estimation of the photon number as long as the total oscillation amplitude is extremely small. We recall that this is a consequence of the fact that we have obtained a dynamical mirror implemented by a harmonic oscillator by assuming that the total deviation from the length $L$ of the cavity is small.

The cavity mode k is initially prepared in the coherent state $\rho_k=\lvert\mu_k\rangle\langle\mu_k\rvert$ and we consider a nonzero external drive. The total amount of photons to lowest order is
\begin{align}
N_k^{(0)}(t)=&(\mu_k+\Lambda_{pk}(t))^2+\left(\Lambda_{xk}(t)\right)^2,
\end{align}
where we recognize the contribution from both the initial amount of photons $\mu_k^2$ in the coherent state, the direct excitation of the cavity mode due to the external drive $\lambda_{xk}(t)$ and $\lambda_{pk}(t)$, and their interplay. 

\subsubsection{First order: $N_k^{(1)}(t)$}\label{N:first:order}
The motion of the wall stimulates the reaction of the field, which leads to the modification of the photon number at the first order in $\epsilon$. Here we estimate the correction of the photon number determined by the dynamics of the mirror. In order to simplify the expressions and facilitate the comparison with our results in Section~\ref{section:second:x}, we switch off all cavity drives as well as the coherent parameter $\mu_{k'}$ in Equation~\eqref{initial:state:complete}, and present the photon number in case of degenerate resonance. We stress that off-resonant terms are smaller than the resonant ones by factors $\omega_k$ or $\omega_k^2$, and can hence be neglected for cavities with large frequencies. The full expression of the first order correction with drives and the coherent excitation of two cavity modes can be found in Appendix \ref{N:first:order:correction}.
We have
\begin{align}
N^{(1)}_k(t)\simeq -\mu_k^2\omega_k t\left(2\lvert\beta\rvert\sin\theta+g\right),
\label{n:photons:first}
\end{align}
where the mechanical drive is again represented by Equation~\eqref{exdr}. 
This formula expresses the resonant exchange of excitations between the mechanical degree of freedom and the mode of the cavity field $k$ when such resonant cavity mode is initially found in a coherent state. Hence, this correction would vanish if we switched the coherent parameter $\mu_k$ off.

We can immediately recognise the contribution of the initial phononic state and the action of the drive. 
The former strictly depends on the coherent parameter suggesting that neither the thermal fluctuation nor the squeezing play a concrete role at the first order.
We notice that it vanishes whenever $\beta$ is a real number, namely when $\theta=n\pi$ with $n\in\mathbb{Z}$, whereas it grows linearly whenever $\beta$ is purely imaginary ($\theta=\pm\pi/2$). 
The second term depends on the amplitude of the external drive and can take positive or negative values.

A direct comparison of this result with both Equation~\eqref{x:first:resonance2} and Equation~\eqref{x:first:resonance3} suggests that the gradual enhancement of the oscillation amplitude due to the degenerate resonance condition takes place without exchange of excitations between the cavity mode $k$ and the mechanical degree of freedom as soon as  $g=-2\sin\theta\lvert\beta\rvert$; whereas it occurs with exchange of excitations if $\beta$ is a pure complex number, with $g\neq-2\sin\theta\lvert\beta\rvert$. Moreover, the efficiency of the exchange of excitations is maximized if $\Im\{\beta\}=g$. This behaviour is confirmed by the investigation of the phonon number, which in case of degenerate resonance reads $N_b(t)=N_b^{(0)}(t)+\epsilon \tilde{N_b}^{(1)}(t)$, with
\begin{align}
N_b^{(0)}(t)\simeq& N_b(0)+g\lvert\beta\rvert\sin\theta+\frac{g^2}{4}\label{n:phonons:zero}\\
\tilde N_b^{(1)}(t)\simeq&\frac{\mu_{k}^2\omega_k t}{2}(2\lvert\beta\rvert\sin\theta+g),
\label{n:phonons:first}
\end{align}
where the initial number of phonons is given by Equation~\eqref{nb0}.

We stress that the direction of conversion (phonons ``down-converted'' into photon pairs or photon pairs ``up-converted'' into phonons) strongly depends on the interplay between the initial phononic coherent state and the action of mechanical drive. This is in contrast with what we will see soon in the next section, by discussing the second order correction by switching off the coherent parameter $\mu_k$. Interestingly, by switching off the mechanical drive it is immediate to obtain the relation
\begin{align}
\langle N_k\rangle_\tau+2\langle N_b\rangle_\tau\simeq& N_k(0)+2N_b(0),
\label{time:average:first:order:relation:betwqeen:number}
\end{align}
which expresses the excitations conservation throughout the dynamics, expected in a lossless system without time-dependent external inputs.

We finally take advantage of our multimode model and compute the number of photons for the mode $k$ in case of non-degenerate resonance between the mechanical mode and two different modes $k$ and $k'$ of the cavity field both initially found in a coherent state with parameter $\mu_k$ and $\mu_{k'}$ respectively. We have
\begin{align}
N^{(1)}_k(t)\simeq & (-1)^{1+k+k'}\mu_k\mu_{k'}\sqrt{\omega_k\omega_{k'}}t\left(2\lvert\beta\rvert\sin\theta+g\right).
\end{align}
It is evident that this expression is the two-mode equivalent of Equation~\eqref{n:photons:first}.

\subsubsection{Second order: $N_k^{(2)}(t)$}\label{N:second:order}
Finally, we conclude the analysis of the photon number by computing the relevant contribution to the number of photons at second order in $\epsilon$ when the field is initially found in its vacuum state. In this case, $N^{(2)}_k(t)=0$ and we note that the first non-zero contribution occurs at second order. Therefore, the existence of such photons is a pure quantum effect.

The second order contribution $N^{(2)}_k(t):=\frac{\omega_k^2 t^2}{2}\Delta N^{(2)}_k(t)$ to the number of photons reads
\begin{align}
N^{(2)}_k(t)=&\frac{\omega_k^2 t^2}{2}\left(\lvert\beta\rvert^2\Delta N^{(2)}_{\beta,k}(t)+\sinh r\, \Delta N^{(2)}_{\textrm{sq},k}(t)\right.\nonumber\\
&+N_T \Delta N^{(2)}_{T,k}(t)+N_T \sinh r \Delta N^{(2)}_{\textrm{sq},T,k}(t)\nonumber\\
&\left.+\Delta N^{(2)}_{\textrm{vac},k}(t)\right)+N_{\textrm{md},k}^{(2)}(t),
\label{N:second:order:tot}
\end{align}
where we distinguished the single contributions according to the physical origin. Here we want to give a physical interpretation of the various elements of this formula, and we leave the explicit form of these terms to Appendix~\ref{N:second:order:correction}. It is crucial to note that \eqref{N:second:order:tot} is exact and does not assume any particular resonant regime. Therefore, all time dependent quantities on the right hand side have nontrivial expressions that, in general, oscillate in time.

The first contribution is $\Delta N^{(2)}_{\beta,k}(t)$, where the index $\beta$ emphasizes that this term stems from the coherent state of the mechanical drive. We notice that this term reproduces the dynamical Casimir effect induced by the harmonic oscillation of the wall. In order to recover the photon number expected by the DCE, we impose the degenerate resonant condition and neglect any off-resonant terms, thereby obtaining and expression for the contribution $N^{(2)}_{\beta,k}(t):=\frac{\omega_k^2 t^2}{2}\Delta N^{(2)}_{\beta,k}(t)$, which reads
\begin{align}
N^{(2)}_{\beta,k}(t)\simeq&\lvert\beta\rvert^2\omega_k^2t^2.
\label{NDCE}
\end{align}
This directly shows the connection between the increasing photon number and the amount of phonons in the coherent state, a connection previously highlighted in the literature \cite{PhysRevA.100.062516,  PhysRevA.64.013808}. Since the motion of the wall only depends on the displacement of the mechanical degree of freedom apart from the external drive, see  Equation~\eqref{x:second:resonance}, we claim that the pair production strictly stems from the ``classical'' oscillation of the wall as long as $\lvert\beta\rvert\gg 1$. This is exactly the regime that reproduces the conditions necessary for the DCE to occur.

The next three terms in bracket, namely $\Delta N^{(2)}_{\textrm{sq},k}(t)$, $\Delta N^{(2)}_{T,k}(t)$ and $\Delta N^{(2)}_{\textrm{sq},T,k}(t)$, are determined by the initial phononic state, in particular by its thermal and squeezing features. We impose the degenerate resonant condition and compute the total contribution $N^{(2)}_{\textrm{sq},T,k}(t):=\frac{\omega_k^2 t^2}{2}(\Delta N^{(2)}_{\textrm{sq},k}(t)+\Delta N^{(2)}_{T,k}(t)+\Delta N^{(2)}_{\textrm{sq},T,k}(t))$, which reads
\begin{align}
N^{(2)}_{\textrm{sq},T,k}(t)\simeq&\omega_k^2t^2\left(\sinh^2r+N_T+2N_T\sinh^2r\right).
\label{NsT}
\end{align}
Interestingly, since the average motion of the mirror neither depends on squeezing of the mechanical state, nor on its thermal fluctuations, the presence of such contributions in \eqref{NsT} and therefore in \eqref{N:second:order:tot} means that both the thermal fluctuations of the phonons and the squeezing of the mechanical mode induce an excitation of the resonant cavity mode, aside from the motion of the mirror ascribable to a classical harmonic oscillation.

We want to stress that all contributions analysed so far arise as a direct consequence of the specific initial phononic state. The term of the interaction Hamiltonian causing the resonant pair creation, proportional to $\hat a^{\dag 2}\hat b$, highlights the excitation transfer between the resonant cavity mode and the harmonic oscillator. This motivates us the compute the second order contribution to the number $N_b(t)$ of phonons at time $t$, which reads
\begin{align}
\Delta N_b^{(2)}(t)=-N_b(0) \frac{\omega_k^2t^2}{2},
\end{align}
and we find once more that, as long as the mechanical drive is switched off, the excitations conservation expressed by Equation~\eqref{time:average:first:order:relation:betwqeen:number} holds.
This means that the increasing number of photons in Equation~\eqref{NDCE} and \eqref{NsT} corresponds to a gradual reduction of the number of phonons. This is exactly what we would expect if the DCE was produced not by an external drive of the wall, but by an initial state of the harmonic oscillator different from the vacuum state.


The term $\Delta N^{(2)}_{\textrm{vac},k}(t)$ in \eqref{N:second:order:tot} stems from the quantization of the mechanical motion of the wall. Although its contribution is negligible at resonance, its existence is consistent with our quantum description of the mechanical motion of the wall, and it is a direct consequence of vacuum fluctuations.

The last term $N_{\textrm{md},k}^{(2)}(t)$ of Equation~\eqref{N:second:order:tot} originates from the mechanical drive. By assuming it in Equation~\eqref{exdr} and imposing the degenerate resonance, we can compute the number of photons $N_{\textrm{md},k}^{(2)}(t)$ that arises from the action of the mechanical drive:
\begin{align}
N_{\textrm{md},k}^{(2)}(t)\simeq&\frac{g\omega_k^2 t^2}{4}(4\lvert\beta\rvert\sin\theta+g).
\label{Nmd}
\end{align}
The first contribution explicitly depends on the real part of the coherent parameter, whereas the second one expresses the possibility to generate photons by forcing the motion of the mirror with the only use of an external drive at resonance with the cavity mode. Note that, if $\beta=0$, photons are created in the cavity, but they do not come at the expense of the depletion of the initial phonons, rather they derive from phonons generated at time $t$ by the mechanical drive itself. This amount of phonons equals the last term in Equation~\eqref{n:phonons:zero}. Along with $N^{(2)}_{\beta,k}(t)$, this term gives rise to the standard DCE following a periodical oscillation of the mirror.

The total number of photons, in the case of degenerate resonance, is achieved by collecting and time-averaging all contributions computed above. We find
\begin{align}
\langle N_k^{(2)}\rangle_\tau\simeq&\frac{\omega_k^2 \tau^2}{3}\left(\lvert\beta\rvert^2\cos^2\theta+\sinh^2r+N_T+2N_T\sinh^2r\right)\nonumber\\
&+\frac{\omega_k^2 \tau^2}{12}\left(g+2 \lvert\beta\rvert\sin\theta\right)^2,
\label{N:k:tot}
\end{align}
where we recovered the explicit interplay between the oscillation induced by the presence of coherent phonons and the motion forced by the external drive that already emerged in Equation~\eqref{x:second:resonance}. As expected, the gradual reduction of the oscillation amplitude, predicted by taking into account the correction on the position of the wall at the third order in $\epsilon$, is brought about by the resonant conversion of phonons into photons. Such phonons can already be present in the initial phonon state, or generated by the external drive, as seen in Equation~\eqref{n:phonons:zero}.

Since the contribution in Equation~\eqref{Nmd} depends on the coherent phase $\theta$, we report two specific scenarios.

\textit{Purely real mechanical coherent parameter.} Whenever $\beta$ is a real number, namely for $\theta=n\pi$ and $n\in\mathbb{N}$, the output number of photons in Equation~\eqref{N:k:tot} reads
\begin{align}
\langle N_k^{(2)}\rangle_\tau\simeq&\frac{\omega_k^2 \tau^2}{3}\left(N_b(0)+\frac{g^2}{4}\right).
\label{N:k:real}
\end{align}
This formula describes the dynamical Casimir effect when the quadratic contribution in $\omega_k$ of the initial phonon number and the action of the external drive interfere constructively with each other.

\textit{Purely imaginary mechanical coherent parameter.}
If $\beta$ is a imaginary quantity, $\theta=(n+1/2)\pi$ with $n\in\mathbb{Z}$, the photon number reduces to
\begin{align}
\langle N_k^{(2)}\rangle_\tau\simeq&\frac{\omega_k^2 \tau^2}{3}\left(\sinh^2r+N_T+2N_T\sinh^2r\right)\nonumber\\
&+\frac{\omega_k^2 \tau^2}{12}\left(g+2(-1)^n \lvert\beta\rvert\right)^2,
\label{N:k:imag}
\end{align}
 
In contrast to the previous case, our results in Equation~\eqref{x:second:resonance3} and Equation~\eqref{N:k:imag} suggest that we can inhibit any contribution of the phononic displacement to the dynamics of the wall, as well as to the resonant generation of photons, by properly tailoring $g$, thereby letting photons arise from both the phononic thermal fluctuation and the intensity of the mechanical squeezing.

We conclude this section by printing the number of photons that are found when the mechanical frequency is non-degenerate resonant with two \textit{different} modes $k$ and $k'$ of the cavity field:
\begin{align}
\langle N_k^{(2)}\rangle_\tau\simeq&\bar N_b(\theta)\frac{\omega_k\omega_{k'}\tau^2}{3},
\end{align}
where we have defined $\bar N_b(\theta):=N_b(0)+g^2/4+g\lvert\beta\rvert\sin\theta$.
\subsection{Force between the mirrors}
The resonance between the cavity mode $k$ and the mechanical oscillation of the wall induces an excitation in the mode, leading to generation of photons, as we have shown above. We expect that such increment of the photon number directly affects the radiation pressure within the cavity acting on the wall, which coincides with the force between the two mirrors in a (1+1)-dimensional scenario. This force can immediately be defined from first principles \cite{Aspelmeyer:Kippenberg:2014}, and it reads $\hat F\equiv -\frac{d\hat H(t)}{dL}$. In our case, it has the expression
\begin{align}
\hat F=&\frac{2\hbar }{L}\sum_{n,m}(-1)^{n+m}\sqrt{\omega_n\omega_m}\,\hat X_n\hat X_{m}-\frac{\lambda_{xb}(t)}{\delta L_0},
\end{align}
where we used the fact that $\epsilon=\delta L_0/L$. For completeness, we restored the vacuum energy of the Hamiltonian to obtain this expression.
 
We want to estimate the average value of the force when the mode of the oscillating mirror is resonant with the mode $k$ of the field. In this case, we find
\begin{align}
\langle F\rangle_\tau\simeq &F_{\textrm{vac}}+\bar N_b(\theta)\frac{\epsilon^2\hbar\omega_k^3\tau^2}{6L},
    \label{F1}
\end{align}
where $F_{\textrm{vac}}:=\sum_{n}\frac{\hbar\omega_n}{2L}$ is the (formally divergent) vacuum contribution and, as usual, we neglect all off-resonant terms.
Clearly, the normal ordering would have removed the presence of $F_{\textrm{vac}}$. However, since we are investigating the time evolution of the radiation pressure in a cavity with no photons at $t=0$, we want to take advantage of our multimode description in order to include the zero-point fluctuations.

We will adopt the reasoning used in the literature \cite{PLUNIEN198687} in order to reinterpret the divergent term in Equation~\eqref{F1}. As a first step, we suppose to have a third (static) mirror located on the right hand side of the movable mirror, say at $L_2$ with $L<L_2$. We can therefore employ the same model used so far in order to describe the field in the region $L<x<L_2$. In particular, the force acting on the movable wall due to the presence of this second cavity is 
\begin{align}
F_2(t)=&F_{\textrm{vac},2}-\sum_{n}\frac{\hbar\omega_n'}{2(L_2-L)}-F_{\textrm{osc},2}(t),
\label{F2}
\end{align}
where $\omega_n'=n\pi c/(L_2-L)$, we have introduced the vacuum fluctuation contribution $F_{\textrm{vac},2}:=-\sum_{n}\frac{\hbar\omega_n'}{2(L_2-L)}$ and we have introduce the function $F_{\textrm{osc},2}(t):=\sum_n\frac{\hbar\delta L_0^2\omega_n'\varphi_1(t)}{(L_2-L)^2}+\sum_{\substack{n,m\\n\neq m}}\frac{\hbar\delta L_0^2\sqrt{\omega_n'\omega_m'}\varphi_2(t)}{(L_2-L)^2}$. The term $F_{2\textrm{osc}}(t)$ includes all time dependent oscillating terms via the functions $\varphi_1(t)$ and $\varphi_2(t)$. These oscillating terms are proportional to $\delta L_0^2$, and therefore they are negligible off-resonance. The function $F_{\textrm{osc},2}(t)$ would given us a term identical (with opposite sign) to the second one of Equation~\eqref{F1} if $L_2=2L$. However, we want to take the limit $L_2\rightarrow\infty$, and therefore we need to keep $L_2$ as a free variable. Therefore we can safely neglect~$F_{\textrm{osc},2}(t)$. 

As a second step, we impose a cut-off of the form $e^{-\gamma\omega_n}$ in the divergent terms of \eqref{F1} and \eqref{F2}. This is a standard procedure and it allows us to neglect any contribution from high frequency modes that cause divergences.
Since $\sum_n n e^{-\alpha n}=e^{\alpha}/(e^{\alpha}-1)^2$, we assume that $\gamma\pi c/L\ll1$ in order to include as many modes as we can. We therefore can expand the divergent terms with cutoff in \eqref{F1} and \eqref{F2} and obtain
\begin{align}
F_{\textrm{vac}}\simeq& \frac{\hbar c}{2\pi\gamma^2}-\frac{\hbar\pi c}{24L^2},\nonumber\\
F_{\textrm{vac},2}\simeq& -\frac{\hbar c}{2\pi\gamma^2}+\frac{\hbar\pi c}{24(L-L_2)^2}.
\label{exp1}
\end{align}
The total force acting on the movable wall is clearly the sum of the two contributions $F_{\textrm{vac}}$ and $F_{\textrm{vac},2}$. We can now safely remove the cutoff $\gamma$ by setting $\gamma=0$, and we can remove the wall inserted at $L_2$ by taking the limit $L_2\rightarrow\infty$. We finally have
\begin{align}
\langle F\rangle_\tau\simeq&-\frac{\hbar\pi c}{24L^2}+\bar N_b(\theta)\frac{\epsilon^2\hbar\omega_k^3\tau^2}{6L},
\label{Casimir:Casimir}
\end{align}

The first term of this expression is the well-known Casimir force between two perfectly conducting plates \cite{PLUNIEN198687}. This force can be interpreted as arising from the zero-point fluctuations of the field inside the cavity, and therefore it holds also at $t=0$. On the contrary, we attribute the remaining positive term to the repulsive force that arises as a consequence of the increasing radiation pressure that forms inside the cavity. Since initially the cavity modes do not contain excitations, this contribution is a direct consequence of the resonant amplification of the field mode $k$. 
In other words, Equation~\eqref{Casimir:Casimir} describes the interplay between static and dynamical Casimir effect, and it highlights the dependence of the force on both the initial phononic state and the intensity of the external drive.



\section{Applications and extensions}\label{appext}
We now briefly discuss potential applications and extensions to our work.

\subsection{Prospective experimental platforms}
We expect that observing the predicted effects in an experimental system will be challenging as the different resonance scenarios i) to iii) in Section~\ref{section:second:x} require a mechanical mode with high frequency exceeding the optical mode frequencies -- i) and ii) -- or matching their difference -- iii). Whilst i) and ii) are only realistic in microwave photonic systems, the latter poses a more modest constraint on the involved modes. For a Fabry-P\'{e}rot type optical resonator this condition can be met, if the mechanical resonator interacts with modes of a long optical cavity of length $L$ that are spaced by the free spectral range $\nu_\text{FSR}=c/2L$. 

\begin{figure}[ht!]
	\centering
	\includegraphics[scale=0.9]{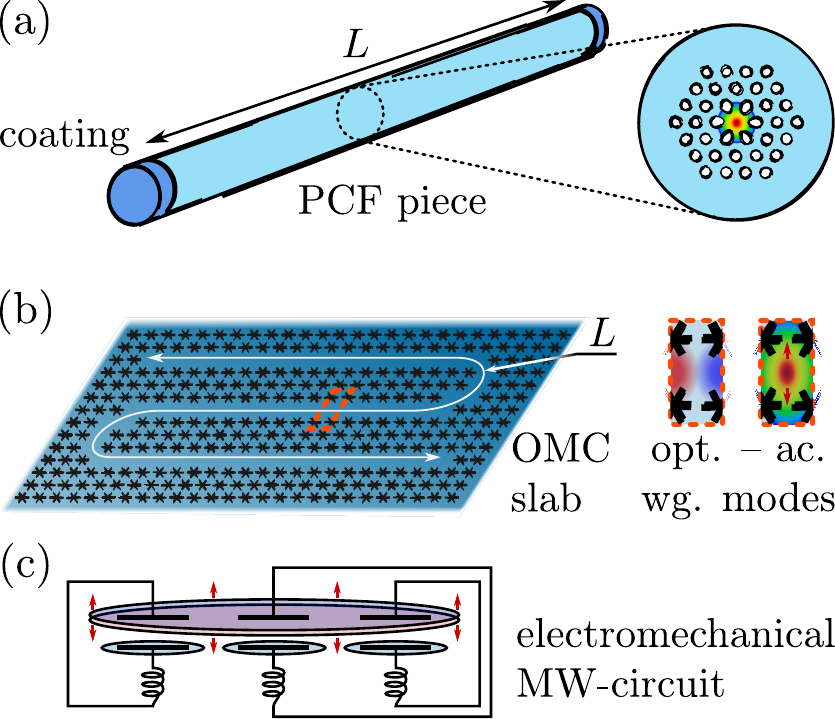}
	\caption{Potential experimental platforms that allow for intermode coupling, where a mechanical resonance frequency matches the optical mode spacing. A photonic crystal fiber (PCF) piece with reflectively coated fiber ends as depicted in (a) can support optical modes with a spacing $c/2L$ that matches the frequency of acoustic breathing modes of the core ($\sim \SI{}{GHz}$) of the core for $L\sim\SI{0.1}{m}$. (b) shows a similar scenario for a optomechanical crystal platform realization, where an extended waveguide of length $L$ supports both optical and acoustic waveguide modes. In (c) an electromechanical platform is depicted featuring a single mechanical resonance coupled to multiple microwave (MW) circuits. The wide range of available MW resonance frequencies would also allow the realizations the degenerate or nondegenerate resonance, however, with an experimentally more limited number of MW resonators.}
	\label{fig:implementations}
\end{figure}

A drawback of this approach is the reduced optomechanical interaction strength that scales with $1/L$ for systems with mechanical resonators that only locally interact with the involved optical modes. Therefore, platforms that make use of extended mechanical resonances as reported in photonic crystal fibers (PCFs) \cite{kang2010all,kang2011reconfigurable,butsch2012optomechanical} or optomechanical crystals (OMCs) \cite{eichenfield2009optomechanical} are more promising. As still evenly spaced optical modes are required, the Fabry-P\'{e}rot type approach should nevertheless be maintained. For PCFs this can be realized by employing a reflective coating at the end facets of a fiber piece \cite{jia2020photonic} of length $L$ as shown in Figure~\ref{fig:implementations}~(a). For a GHz range mechanical mode \cite{kang2010all,kang2011reconfigurable} the required length will be on the order of $\SI{0.1}{\meter}$ to realize evenly spaced optical modes within a low dispersion region of the guided mode spectrum. This length will be reduced for OMCs as silicon exhibits a higher refractive index and as higher frequency acoustic modes are realized \cite{ren2020two}. Still, up to millimeter long waveguide paths that support both optical and acoustic modes with photonic and phononic reflectors at the ends will be required \cite{fang2016optical,fang2017generalized} for example using two-dimensional OMC realizations \cite{safavi2014two,ren2020two} as sketched in Figure~\ref{fig:implementations}~(b). For a close to ideal mode overlap of a low-dispersion optical and a $\Gamma$-point mechanical mode, the resulting vacuum optomechanical coupling strength $g_0$ scales with the number of required unit cells $n_\text{uc}$ as $1/\sqrt{n_\text{uc}}$ as the effective mass of the mechanical mode linearly increases with $n_\text{uc}$. For $n_\text{uc}\sim 10^3$ and a waveguide optomechanical coupling of $\approx\SI{1.5}{\mega\hertz}$ \cite{ren2020two}, this would result in $g_0\approx\SI{50}{\kilo\hertz}$. Such approaches, especially if slight optimizations are conducted, will therefore result in experimentally accessible systems for resonance scenario iii). 

Electromechanical systems using superconducting microwave resonators \cite{wilson2011observation, RevModPhys.84.1} with a displacement dependent capacitance provide experimental platforms with smaller photonic mode frequencies. With using higher frequency mechanical modes this would allow realizations of the other resonance scenarios. The limitation however will be the number of microwave modes coupled to the same mechanical element meeting the resonance condition. Several tuned and simultaneously coupled microwave resonators will be required as indicated in Figure~\ref{fig:implementations}~(c). This strongly restricts the number of involved modes in contrast to Fabry-P\'{e}rot type photonic resonators, where a large number of evenly spaced optical modes naturally occurs. 

Whilst not giving the direct interpretation as a classical dynamic Casimir effect as in a system of two opposing mirrors, analogous electro- or optomechanical systems support experimentally accessible regimes that can realize all three resonance scenarios. The dynamics of involved photonic modes would be reproduced by such systems providing means for single mode squeezing, two mode squeezing or mode mixing applications. 

\subsection{Scalar massive field}
The procedure presented in Section \ref{theoretical:model} can straightforwardly be extended to the case of confined massive field. Let us suppose that a massive real scalar field $\phi(t,x)$ is confined in a static cavity. The classical Lagrangian density $\mathcal{L}(t,x)$ in $1+1$ dimensions reads
\begin{align}\label{lagrangian:density}
\mathcal{L}=\frac{\hbar}{2}\left(\frac{\partial^2\phi}{c^2\partial t^2}-\frac{\partial^2\phi}{\partial x^2}-\frac{M^2c^2}{\hbar^2}\phi^2\right),
\end{align}
where $M$ is the mass.
The classical field is solution to the (1+1)-dimensional Klein Gordon equation for a massive field that reads $(\square-M^2c^2/\hbar^2)\phi(t,x)=0$.
We again make the ansatz $\phi_n(t,x)=e^{-i\omega_n t}\sin(p_n x)$, where the dispersion relation now reads $\omega_n=(\hbar^2 c^2 p_n^2+M^2c^4)^{\frac{1}{2}}$, with wave vector $p_n:=\frac{n\pi}{L}$. The explicit expression of both the field $\phi(t,x)$ and its conjugate momentum $\Pi(t,x)$ do not differ from those found before, except for the massive term in the frequencies $\omega_n$. 

We can therefore apply the same procedure and find the classical density Hamiltonian of the massive field, which reads $\mathcal{H}=\frac{1}{2}\left(\Pi^2(t,x)+(\partial_x\phi(t,x))^2+\frac{M^2c^2}{\hbar^2}\phi^2(t,x)\right)$. This, in turn, allows us to compute the Hamiltonian operator $\hat{H}=\hat H_0+\epsilon\hat H_{\textrm{I}}$, where
\begin{align}\label{quantum:hamiltonian:massive:terms2}
\hat{H}_0:=&\sum_{n}\hbar\omega_n\,\hat{a}_n^\dag\hat{a}_n+\hbar\omega\hat{b}^\dag\hat{b}\nonumber,\\
\hat{H}_{\textrm{I}}:=&-2\sum_{n}\frac{\hbar c^2 p_n^2}{\omega_n}\,\hat{a}_n^\dag\hat{a}_n \hat X_{b}-\sum_{n}\frac{\hbar c^2 p_n^2}{\omega_n}\,\left(\hat{a}_n^{\dag2}+\hat{a}_n^2\right)\hat X_{b}\nonumber\\
&-4\sum_{n\neq m}(-1)^{n+m}\frac{\hbar c^2 p_n p_m}{\sqrt{\omega_n\omega_m}}\,\hat X_n\hat X_{m}\hat X_{b}.
\end{align}
We notice that $\hat H_0$ is formally identical to the free Hamiltonian of the massless case, whereas the interaction Hamiltonian $\hat H_{\textrm{I}}$ differs from Equation~\eqref{quantum:hamiltonian:terms2} only in the coupling constant, which now includes the dispersion relation for the massive field. For this reason, the extension of our previous results to the massive scenario can be performed straightforwardly. As a single example, we report the number of massive excitations generated at time $t$ in case of degenerate resonance with a cavity mode $k$, when the input state is given by $\hat{\rho}(0)=\prod_{n}\lvert 0_n\rangle\langle 0_n\rvert\otimes\hat{\rho}_{\textrm{m}}^{\textrm{(DST)}}$ in the vacuum state of the massive field. To second order in $\epsilon$ we obtain
\begin{align}
N_k^{(2)}(t)\simeq&\frac{\epsilon^2 c^4 p_k^4 t^2}{\omega_k^2}N_b(0).
\end{align}

\section{Considerations and outlook}\label{consout}

\subsection{Considerations}
We now proceed with some considerations regarding our work.

First, we notice that the combination $\epsilon\omega_k$ in our Hamiltonian exactly corresponds to the vacuum optomechanical coupling strength \cite{Aspelmeyer:Kippenberg:2014}. This allows us to interpret $\delta L_0$ as the zero-point fluctuation of the quantum harmonic oscillator, namely $\delta L_0\equiv x_{ZPF}=\sqrt{\hbar/(2M\omega)}$ (with $M$ mass of the mirror), whose order of magnitude is typically much less than the length of the whole cavity \cite{doi:10.1126/science.1156032}. This argument reinforces the perturbative approach employed in this work. 

\begin{figure*}
	\centering
	\includegraphics[width=1\linewidth]{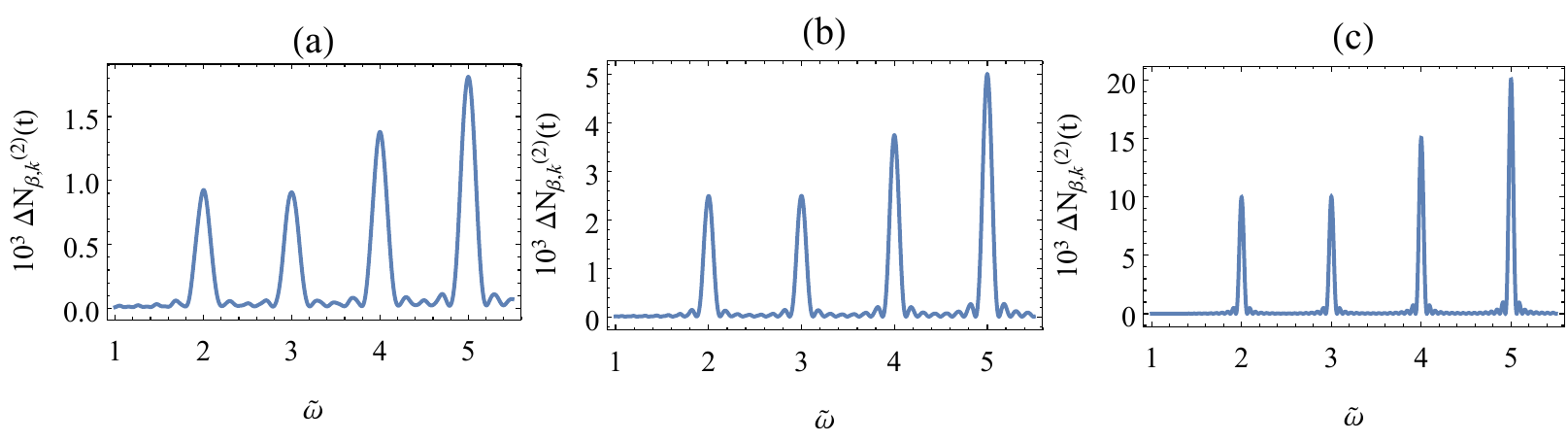}
	\caption{Second order correction $\Delta N_{\beta,k}^{(2)}(t)$ of the average number of excitations in mode $k$ as a function of the dimensionless oscillation frequency $\tilde{\omega}:=\frac{L}{\pi c}\omega$ of the wall at different dimensionless times $\tilde{t}:=\frac{\pi c}{L}t$. We have plotted: (a) at time $\tilde t=30$, (b) at time $\tilde t=50$, and (c) at time $\tilde t=100$. In all figures, the four peaks correspond to the resonant frequency at $\tilde\omega=2\tilde\omega_1=2$, $\tilde\omega=\tilde\omega_1+\tilde\omega_2=3$, $\tilde\omega=\tilde\omega_1+\tilde\omega_3=4$ and $\tilde\omega=\tilde\omega_1+\tilde\omega_4=5$. The mass of the mirror is $M=10^{-16}$kg.}
	\label{nk}
\end{figure*}
We also want to point out that the inverse of the coupling strength determines the regime of validity of our perturbative approach, as observed in Equation~\eqref{decay}. This formula expresses the reduction of the oscillation amplitude at resonance, and it holds as long as $t\ll t_c\equiv 2/\epsilon\omega_k$.
In typical optomechanical experiments this quantity ranges between $10^{-6}$s-$10^{-2}$s \cite{verhagen2012quantum, teufel2011sideband, chan2011laser, murch2008observation}, but it can dramatically reduce in superconducting quantum-interference devices \cite{PhysRevLett.103.147003, PhysRevA.82.052509}.
On the other hand, the inverse of the oscillation frequency, $1/\omega$, gives an estimation of the scale time for resonant peak in the photon number to emerge. 
Such peak increases its intensity rapidly in time, as shown in Figure \ref{nk}, where the correction $\Delta N_{\beta,k}^{(2)}(t)$ in Equation~\eqref{deltaNbeta} is plotted by scanning the oscillation frequency at different time, where both frequency and time are normalized as following: $\tilde\omega=\frac{L}{\pi c}\omega$ and $\tilde t= \frac{\pi c}{L}t$. 
The graph shows how the sinc functions of Equation~\eqref{deltaNbeta} tend to a Dirac delta function at resonance when $\tilde t\gg 1$, or $t\gg 2/\omega_k$.

\begin{figure*}
	\centering
	\includegraphics[width=1\linewidth]{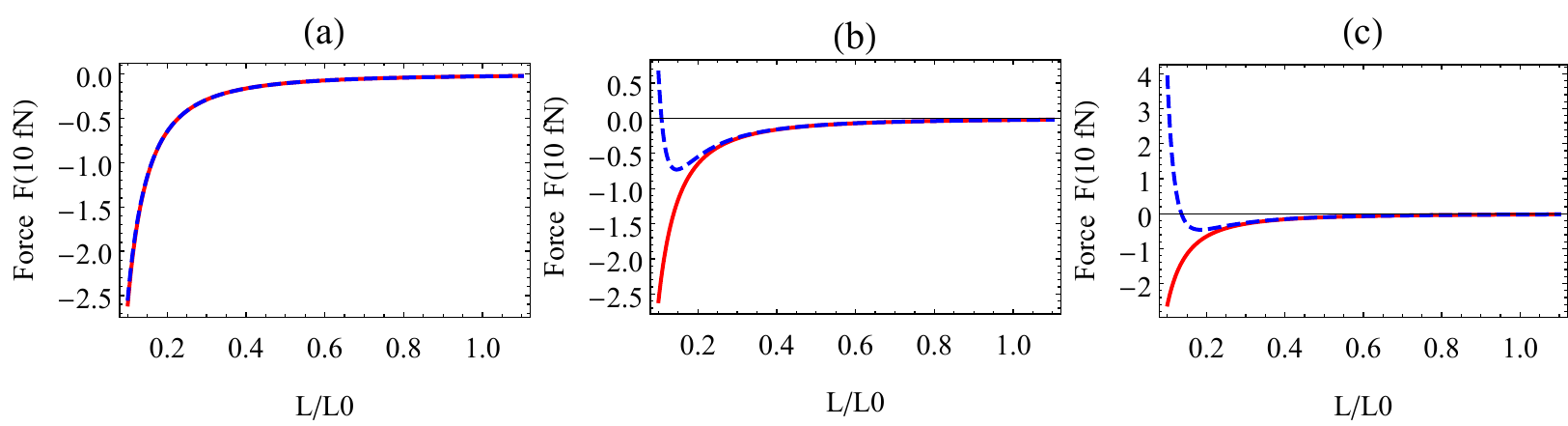}
	\caption{The Casimir force between the two wall of the cavity at two different times, namely $t=0\,s$ (solid red line), and $t=10^{-6}\,s$ (blue dashed line). Each panel has been obtained using different coherent phonon number: (a) $\lvert\beta\rvert^2=1$, (b) $\lvert\beta\rvert^2=50$, and (c) $\lvert\beta\rvert^2=100$. We are working here in the degenerate resonant regime, where the oscillation frequency $\omega$ reads $\omega=2\omega_1$. Other parameters are the following: cavity length $L=10\mu$m, mirror mass $M=10^{-16}$kg.}
	\label{Casimir:force}
\end{figure*}

Now we want to give an estimation of the order of magnitude of the second order correction on the Casimir force.  
The expected dominant repulsive behaviour of the force emerges at $L<L_{c}$, where the \textit{critical length} $L_{c}$ is defined by $\langle F(L_c)\rangle_\tau=0$, namely
\begin{align}
L_{c}=\left(\frac{4\pi \bar N_b(\theta) c\,\hbar \, \tau^2}{M}\right)^{\frac{1}{3}}.
\end{align}
Moreover, the new trend of the Casimir force suggests the presence of a minimum $\langle F(L_{min})\rangle_\tau$ that is reached at $L_{min}=(5/2)^{1/3}L_c$ and has the expression
\begin{align}
\langle F(L_{min})\rangle_\tau\simeq&-\frac{\hbar\pi c}{40L_{min}^2}.
\label{Casimir:Casimir2}
\end{align}
In order to provide a concrete evaluation of these quantities, we switch off the external drive, fix $T=0$ and $r=0$. Moreover, we fix a reference length of the cavity when the wall is at rest $L_0=10\mu$m \cite{PhysRevLett.111.060403}, and assume that the oscillating mirror has mass $M\approx 10^{-16}$kg. 
We want to resonantly excite the lowest mode of the cavity, namely $\omega=2\omega_1$. As can be seen in Figure~\eqref{Casimir:force}(a), where we plotted the trend of the Casimir force by varying the distance between the two walls at different time, no inversion of the sign of the force is observed. This suggests that correction obtained above to the standard Casimir force does not play an appreciable role when $t\ll t_c$, namely where the radiation pressure is weak due to the low amount of generated photons. On the contrary, the repulsive trend becomes appreciable with the increase of the phonon number, as shown in Figure~\ref{Casimir:force}(b) and (c), where the inversion point is visible for short distances. In particular, this amounts to $L_c/L_0\simeq 0.11$ in Figure~\ref{Casimir:force}(b) and to $L_c/L_0\simeq 0.14$ in Figure \ref{Casimir:force}(c). Furthermore, the total number of photons at $t=10^{-6}$ ranges from $N_k\approx 10^{-6}$ in Figure \ref{Casimir:force}(a) up to $N_k\approx 10^{-4}$ in Figure \ref{Casimir:force}(c).

One important aspect of the the model proposed above is that it allows us to compute the excitation transfer between the wall and the field.
As already known \cite{PhysRevA.100.062516}, this is carried out by the squeezing term in the interaction Hamiltonian proportional to $(\hat a_k^{\dag})^2\hat b$, and can occur when the phonon state is initially found in an excited state. However, we want to emphasize that our model predicts the resonant excitation of the cavity mode by means of an external drive acting on the phononic degree of freedom. In particular, the term $(g\omega_k t/2)^2$ in Equation~\eqref{Nmd} stems from a time-dependent displacement term in the Hamiltonian, acting on the quantized mechanical mode.

The specific choice of the external drive in Equation~\eqref{exdr} gives rise to a smooth transition from a static to a dynamical regime, which terminates once the sinusoidal motion of the wall is stabilized. A similar equation of motion has been previously employed \cite{PhysRevA.64.013808}, where the dynamics is not included in the Hamiltonian but enters the mathematical description of the system in terms of time-dependent Dirichlet boundary conditions of the equation of motion. In our model, the use of a time-dependent displacement acting on the mechanical degree of freedom avoids the requirement of solving differential equations with time-dependent boundary conditions.  As expected, we can recover both the equation of motion and the number of photons in \cite{PhysRevA.64.013808} by imposing either $g=1$ and $\beta=0$ or $g=0$ and $\beta=\frac{1}{2}e^{i\pi/2}$.

In addition, by assuming $\omega_k^2t^2\lvert\beta\rvert^2=\omega_k^2t^2g^2/4=N_{0}/2$, $r=0$ and $T=0$ in Equation~\eqref{N:k:tot}, our model predicts an interference of the photon number of the form $N_k=N_{0}(1+\sin\theta)$, caused by the interplay between coherent phonons and mechanical drive, whose pattern depends on the coherent phase $\theta$. In particular, an effect of amplification occurs at $\theta=\pi/2$, whereas the inhibition of the output radiation is expected whenever $\theta=-\pi/2$. Coherently, such values correspond to the maximal value of the oscillation amplitude and the arrest of the oscillating wall respectively, see Equation~\eqref{x:oscill}. Since the DCE as a squeezing phenomenon stimulates the creation of photon pairs, and the above-described modulation of $N_0$ involves both photons at the same time, by scanning the phase $\theta$ we can reproduce the interference pattern expected by a SU(1,1) interferometer characterized by two optical parametric amplifiers in the low gain regime \cite{PhysRevA.33.4033, PhysRevA.104.043707, manceau2017improving, Ferreri2021spectrallymultimode}. We leave the study of this aspect to future work.

\subsection{Outlook}
In this work we showed that the quantization of the harmonic oscillation of a cavity wall translates into the presence of a further ``mechanical'' quantum degree of freedom, which in turn  gives rise to both the optomechanical coupling and specific terms determining the exchange of quantum excitations between photons and phonons. More specifically, we have seen that, in a multimode description of the cavity field, the latter degrees of freedom are responsible for effects of multimode squeezing, which can emerge by properly tuning the oscillation frequency of the harmonic oscillator, or its external driving force. 

The description of the whole system that we have here achieved is general enough to be exported to many systems, as well as to be used for modelling of concrete physical systems. Apart from the theoretical and foundational knowledge gained by following the procedure proposed here, we also highlight the potential implications for future research. For example, interesting applications can be found in continuous variables quantum information and computing, where single- and two-mode quantum gates on bosonic degrees of freedom are necessary as ingredients for the achievement of desired tasks \cite{RevModPhys.84.621}. Universal quantum computing also requires nonlinear bosonic gates, i.e., unitary operations that are induced by Hermitian operators that are at least cubic in the creation and annihilation operators (or, equivalently, in the quadrature operators). While single- and two-mode gates have already been discussed in the context of resonances for field modes of cavity fields \cite{Bruschi_2013, PhysRevD.86.105003, PhysRevLett.111.090504}, our current work moves beyond the ideal scenario where abstract boundary conditions are varied as functions of time and allows to concretely study the gate-induced operations in relation to the additional interplay between the different degrees of freedom. While our work does not include decoherence or dissipation, which are are significant and unavoidable in cavity systems \cite{PhysRevA.104.013501, PhysRevX.8.011031}, we believe that the current study provides the first steps towards more concrete realizations. However, we expect that the dynamic effects, as show in Fig.~\ref{nk}, will persist in a full treatment as their time-scales are faster than the dissipation or decoherence of optical or mechanical modes reached in current experimental systems. For this reason, we leave it to future work to include the effects of cavity loss and more realistic cavity designs.

\section{Conclusion}\label{conclusion}
In this paper we have studied the dynamics of a quantum field trapped within a cavity with a moving quantized wall. Our approach tackles the problem at the fundamental Hamiltonian level and allows us to obtain a Hamiltonian that includes the interaction between the field and the degree of freedom of the wall \textit{avoiding} the need for time-dependent (Dirichlet) boundary conditions. The Hamiltonian contains both the standard optomechanical interaction term as well as the terms responsible for field dynamics, thereby allowing us to explore each regime separately, together with their interplay. We achieved this result by adding external driving terms for both the field and mirror-displacement, and by considering specifically tailored initial states of the whole system.

We employed our model to estimate the time evolution of the position of the wall, the average photon number in the cavity, and the pressure that acts on the wall. We focused our efforts on the resonant regime, where either the intrinsic frequency of the wall or the frequency of the external drive are twice the frequency of one of the cavity modes. This regime appears naturally in many physical systems such as, for example, degenerate parametric down-conversion \cite{doi:10.1080/00107514.2018.1488463, PhysRevA.50.5122}, and four-wave mixing \cite{Abrams:78, Yuen:79}. We were able to reproduce the standard optomechanical regime or the dynamical Casimir effect regime by specifically tailoring the initial state, and the external mechanical drive. We also computed the excitation transfer between the cavity modes and the vibrational mode of the wall, which occurs at the interplay between these two scenarios. When the wall is initially displaced but not driven, we found that this excitation exchange translates into the photon pair production via DCE by means of the conversion of phonons into photons, leading to a decrease in amplitude of the wall oscillations. Similarly, the action of an external drive, initially pushing the wall at the resonant frequency with one of the cavity mode, causes the reaction of the field, which again translates into the vacuum squeezing of the resonant mode (DCE).

Furthermore, we took advantage of the multimode structure of the cavity field in order to study the time evolution of the radiation pressure when the cavity field is initially prepared in its vacuum state, therefore obtaining the combined action of the static Casimir force and the repulsive radiation pressure caused by the increase of cavity photons by means of the dynamical Casimir effect. In addition, we extended our model to the case of massive scalar fields, for which the results are formally equivalent to those achieved in the massless scenario and therefore do not imply qualitative differences, and we also proposed a physical system based on a photonic crystal fiber for potential concrete implementation of the techniques that we have developed.
Finally, applications of this model include single- and two-mode quantum gates acting on the field modes in the cavity. These operations can be used as core ingredients for quantum information processing with continuous variables, for example in quantum computing and the realization of multimode quantum gates.

The proposed approach is derived from first principles that follow a general procedure, and it can therefore be used to tackle other applications where a boundary is quantized without the need of solving time-dependent differential equations. 
We believe that this work provides an important addition to the study of the interplay of quantum field theory and optomechanics, as well as novel theoretical tools to be used in the development of future quantum technologies.

\acknowledgments
We thank Roberto Passante and Lucia Rizzuto for interesting
discussions, as well as Franco Nori, Vincenzo Macrì, and
Jorma Louko for the useful comments. This work has received
funding from the European Union’s Horizon 2020 program
under the ERC consolidator grant RYD-QNLO (Grant No. 771417). F.K.W., A.F., and D.E.B. acknowledge the program Geqcos, sponsored by the German Federal Ministry of Education and Research, under the funding program “quantum technologies–from basic research to the market”,
No. 13N15685. H.P. and S.H. acknowledge funding by the
Deutsche Forschungsgemeinschaft (DFG, German Research
Foundation) under Germany’s Excellence Strategy, Cluster
of Excellence Matter and Light for Quantum Computing
(ML4Q) EXC 2004/1, 390534769.

\bibliographystyle{apsrev4-2}
\bibliography{biblio}

\appendix
\onecolumngrid

\newpage

\section{Interaction Hamiltonian}\label{app:tools}
The effective interaction Hamiltonian calculated via Equation~\eqref{H:tilde} is
\begin{align}\label{quantum:hamiltonian:interaction:term3}
\hat{\tilde{H}}_{\textrm{I}}(t)=&-\hbar\epsilon\left[\vartheta(\omega_{k},\omega_{k'},t)+\sum_{n}\frac{\omega_n}{2}\,\left(2\hat{a}_n^\dag\hat{a}_n+e^{2i\omega_n t}\hat a_n^{\dag 2}+e^{-2i\omega_n t}\hat a_n^{2}\right)\right.\nonumber\\
&+\sum_{\substack{n,m\\n\neq m}}(-1)^{n+m}\frac{\sqrt{\omega_n\omega_m}}{2}\left(e^{i\omega_nt}\hat a_n^\dag+e^{-i\omega_nt}\hat a_n\right)\left(e^{i\omega_mt}\hat a_m^\dag+e^{-i\omega_mt}\hat a_m\right)\nonumber\\
&-2\sum_{\substack{n\\n\neq k}}(-1)^{n+k}\sqrt{\omega_n\omega_{k}}\psi_{k}(t)\left(e^{i\omega_n t}\hat a_n^{\dag}+e^{-i\omega_n t}\hat a_n\right)-2\sum_{\substack{n\\n\neq k'}}(-1)^{n+k'}\sqrt{\omega_n\omega_{k'}}\psi_{k'}(t)\left(e^{i\omega_n t}\hat a_n^{\dag}+e^{-i\omega_n t}\hat a_n\right)\nonumber\\
&+i\omega_{k'}\left(e^{2i\omega_{k'}t}\Lambda_{k'}^-(t)\hat a_{k'}^\dag-e^{-2i\omega_{k'}t}\Lambda_{k'}^-(t)\hat a_{k'}-i\omega_{k'}\Lambda_{pk'}(t)(\hat a_{k'}^\dag+\hat a_{k'})-\Lambda_{xk'}(t)(\hat a_{k'}^\dag-\hat a_{k'}) \right)\nonumber\\
&\left.+i\omega_{k}\left(e^{2i\omega_{k}t}\Lambda_{k}^-(t)\hat a_{k}^\dag-e^{-2i\omega_{k}t}\Lambda_{k}^-(t)\hat a_{k}-i\Lambda_{pk}(t)(\hat a_{k}^\dag+\hat a_{k})-\Lambda_{xk}(t)(\hat a_{k}^\dag-\hat a_{k})\right)
\right]\left(e^{i\omega t}\hat b^\dag+e^{-i\omega t}\hat b+2\xi(t)\right).
\end{align}
where we have defined the following auxiliary functions:
\begin{align}
\Lambda_j^-(t)=&\Lambda_{xj}(t)-\Lambda_{pj}(t),\\
\xi(t)=&\cos(\omega t)\Lambda_{pb}(t)-\sin(\omega t)\Lambda_{xb}(t),\\
\psi_2(t)=&\cos(\omega_{k}t)\Lambda_{pk}(t)+\sin(\omega_{k}t)\Lambda_{xk}(t),\\
\psi_3(t)=&\cos(\omega_{k'}t)\Lambda_{pk'}(t)+\sin(\omega_{k'}t)\Lambda_{xk'}(t),\\
\vartheta(\omega_{k},\omega_{k'},t)=&-\omega_k\left([\Lambda_{pk}(t)]^2+[\Lambda_{xk}(t)]^2\right)-2\omega_{k}\sin(2\omega_{k}t)\Lambda_{pk}(t)\Lambda_{xk}(t)-\omega_{k}\cos(2\omega_{k} t)\left([\Lambda_{xk}(t)]^2+[\Lambda_{pk}(t)]^2\right)
\nonumber\\
&-\omega_{k'}\left([\Lambda_{pk'}(t)]^2+[\Lambda_{xk'}(t)]^2\right)-2\omega_{k'}\sin(2\omega_{k'}t)\Lambda_{pk'}(t)\Lambda_{xk'}(t)-\omega_{k'}\cos(2\omega_{k'} t)\left([\Lambda_{xk'}(t)]^2+[\Lambda_{pk'}(t)]^2\right)\nonumber\\
&+(-1)^{k+k'}4\sqrt{\omega_k\omega_{k'}}\left(\sin(\omega_{k}t)\sin(\omega_{k'}t)[\Lambda_{xk'}(t)]^2+\cos(\omega_{k}t)\cos(\omega_{k'}t)[\Lambda_{pk'}(t)]^2\right),\nonumber\\
\end{align}
with $j\equiv k, k'$ or $b$.
\newpage
\section{Position of the wall: second order correction}\label{x:second:order}
The correction to the position of the wall estimated to the second order in $\epsilon$ is performed by calculating the second term of Equation~
\eqref{time:evolution:second:order} when $\hat{A}\equiv \hat X$. This gives:
\begin{align}
\frac{\Delta x^{(2)}(t)}{\delta L_0}=&\frac{\omega_{k}\mu_{k}^2 t}{2}\left[\omega t\sinc^2\left(\frac{\omega t}{2}\right)+\sin\left[\frac{(2\omega_{k}-\omega)t}{2}\right]\sinc\left[\frac{(2\omega_{k}+\omega)t}{2}\right]+\sin\left[\frac{(2\omega_{k}+\omega)t}{2}\right]\sinc\left[\frac{(2\omega_{k}-\omega)t}{2}\right]\right]\nonumber\\
&+\frac{\omega_{k'}\mu_{k'}^2 t}{2}\left[\omega t\sinc^2\left(\frac{\omega t}{2}\right)+\sin\left[\frac{(2\omega_{k'}-\omega)t}{2}\right]\sinc\left[\frac{(2\omega_{k'}+\omega)t}{2}\right]+\sin\left[\frac{(2\omega_{k'}+\omega)t}{2}\right]\sinc\left[\frac{(2\omega_{k'}-\omega)t}{2}\right]\right]\nonumber\\
&-(-1)^{k+k'}\sqrt{\omega_k\omega_{k'}}\mu_k\mu_{k'}t\left[\sin\left[\frac{(\omega_k-\omega_{k'}+\omega)t}{2}\right]\sinc\left[\frac{(\omega_k-\omega_{k'}-\omega)t}{2}\right]\right.\nonumber\\
&+\sin\left[\frac{(\omega_k-\omega_{k'}-\omega)t}{2}\right]\sinc\left[\frac{(\omega_k-\omega_{k'}+\omega)t}{2}\right]+\sin\left[\frac{(\omega_k+\omega_{k'}-\omega)t}{2}\right]\sinc\left[\frac{(\omega_k+\omega_{k'}+\omega)t}{2}\right]\nonumber\\
&\left.+\sin\left[\frac{(\omega_k+\omega_{k'}+\omega)t}{2}\right]\sinc\left[\frac{(\omega_k+\omega_{k'}-\omega)t}{2}\right]\right]\nonumber\\
&+i(-1)^{k+k'}\sqrt{\omega_k\omega_{k'}}\left[\mu_{k'}\left( e^{i\omega t}\int_0^tdt'\psi_{k}(t')e^{i(\omega_{k'}-\omega)t'}-e^{-i\omega t}\int_0^tdt'\psi_{k}(t')e^{-i(\omega_{k'}-\omega)t'}\right.\right.\nonumber\\
&\left.+e^{i\omega t}\int_0^tdt'\psi_{k}(t')e^{i(\omega_{k'}+\omega)t'}-e^{-i\omega t}\int_0^tdt'\psi_{k}(t')e^{-i(\omega_{k'}+\omega)t'}\right)+\mu_{k}\left( e^{i\omega t}\int_0^tdt'\psi_{k'}(t')e^{i(\omega_{k}-\omega)t'}\right.\nonumber\\
&\left.\left.-e^{-i\omega t}\int_0^tdt'\psi_{k'}(t')e^{-i(\omega_{k}-\omega)t'}+e^{i\omega t}\int_0^tdt'\psi_{k'}(t')e^{i(\omega_{k}+\omega)t'}-e^{-i\omega t}\int_0^tdt'\psi_{k'}(t')e^{-i(\omega_{k}+\omega)t'}\right)\right]\nonumber\\
+&\mu_{k}\omega_k\left[i\left(e^{-i\omega t}\int_0^tdt'\Lambda_{pk}(t')e^{i\omega t'}-e^{i\omega t}\int_0^tdt'\Lambda_{pk}(t')e^{-i\omega t'}\right)+\frac{1}{2}\left(e^{-i\omega t}\int_0^tdt'\Lambda_{k}^-(t')e^{i(2\omega_{k}+\omega) t'}\right.\right.\nonumber\\
&\left.\left.+e^{i\omega t}\int_0^tdt'\Lambda_{k}^-(t')e^{i(2\omega_{k}+\omega) t'}-e^{-i\omega t}\int_0^tdt'\Lambda_{k}^-(t')e^{-i(2\omega_{k}-\omega) t'}-e^{i\omega t}\int_0^tdt'\Lambda_{k}^-(t')e^{i(2\omega_{k}-\omega) t'}\right)\right]\nonumber\\
+&\mu_{k'}\omega_{k'}\left[i\left(e^{-i\omega t}\int_0^tdt'\Lambda_{pk'}(t')e^{i\omega t'}-e^{i\omega t}\int_0^tdt'\Lambda_{pk'}(t')e^{-i\omega t'}\right)+\frac{1}{2}\left(e^{-i\omega t}\int_0^tdt'\Lambda_{k'}^-(t')e^{i(2\omega_{k'}+\omega) t'}\right.\right.\nonumber\\
&\left.\left.+e^{i\omega t}\int_0^tdt'\Lambda_{k'}^-(t')e^{i(2\omega_{k'}+\omega) t'}-e^{-i\omega t}\int_0^tdt'\Lambda_{k'}^-(t')e^{-i(2\omega_{k'}-\omega) t'}-e^{i\omega t}\int_0^tdt'\Lambda_{k'}^-(t')e^{i(2\omega_{k'}-\omega) t'}\right)\right]\nonumber\\
&+\frac{1}{2i}\left(e^{i\omega t}\int_0^tdt'\vartheta(\omega_{k},\omega_{k'},t')e^{-i\omega t}-e^{-i\omega t}\int_0^tdt'\vartheta(\omega_{k},\omega_{k'},t')e^{i\omega t} \right).
\label{xcompl}
\end{align}
\newpage
\section{Number of photons: first order correction}\label{N:first:order:correction}
The correction to the photon number in the mode $k$ estimated up to the first order is performed by calculating the second term of Equation~
\eqref{time:evolution:second:order} when $\hat{A}\equiv \hat a_k^\dag\hat a_k$. This reads:
\begin{align}
\Delta N_k^{(1)}(t)=&-2\mu_k^2\omega_k\left[\lvert\beta\rvert t\sin\left[\frac{(2\omega_k+\omega)t-2\theta}{2}\right]\sinc\left[\frac{(2\omega_k+\omega)t}{2}\right]+\lvert\beta\rvert t\sin\left[\frac{(2\omega_k-\omega)t+2\theta}{2}\right]\sinc\left[\frac{(2\omega_k-\omega)t}{2}\right]\right.\nonumber \\
&\left.+2 \int_0^t dt'\xi(t')\sin(2\omega_k t')\right]+(-1)^{k+k'}\sqrt{\omega_k\omega_{k'}}\mu_k\mu_{k'}\lvert \beta\rvert t\left[\frac{4}{\lvert \beta\rvert t}\int_0^t dt' \xi(t')\sin\left(\omega_{k'}t'\right)\cos\left(\omega_k t'\right)\right.\nonumber\\
&+\sin\left[\frac{(\omega_k+\omega_{k'}+\omega)t-2\theta}{2}\right]\sinc\left[\frac{(\omega_k+\omega_{k'}+\omega)t}{2}\right]+\sin\left[\frac{(\omega_k+\omega_{k'}-\omega)t+2\theta}{2}\right]\sinc\left[\frac{(\omega_k+\omega_{k'}-\omega)t}{2}\right]\nonumber\\
&\left.+2\cos\left[\frac{(\omega t-2\theta)}{2}\right]\sin\left[\frac{(\omega_k-\omega_{k'})t}{2}\right]\left(\sinc\left[\frac{(\omega_k-\omega_{k'}+\omega)t}{2}\right]+\sinc\left[\frac{(\omega_k-\omega_{k'}-\omega)t}{2}\right]\right)\right]\nonumber\\
&+4(-1)^{k+k'}\sqrt{\omega_k\omega_{k'}}\mu_k\int_0^t dt'\sin(\omega_k t')\psi_{k'}(t')\left(2\cos(\omega t'-\theta)+\xi(t')\right)\nonumber\\
&+4\omega_k\mu_k\int_0^t dt'\left[\lvert\beta\rvert\Lambda_{xk}(t')\cos(\omega t'-\theta)+\Lambda_{k}^-(t')\cos(2\omega_k t')\cos(\omega t'-\theta)+\Lambda_{k}^-\xi(t')\cos(2\omega_k t')-\Lambda_{xk}(t')\xi(t')\right]\nonumber\\
&-2\Lambda_{xk}(t)\left\{2\mu_k\omega_k t\cos\left[\frac{(\omega t-2\theta)}{2}\right]\sinc\left(\frac{\omega t}{2}\right)+2\mu_k\omega_k\int_0^t dt'\xi(t')+2\mu_k\omega_k\int_0^t dt'\xi(t')\cos(2\omega_kt')\right.\nonumber\\
&+\mu_k\lvert\beta\rvert\omega_k t\cos\left[\frac{(2\omega_k+\omega)t+2\theta}{2}\right]\sinc\left[\frac{(2\omega_k+\omega)t}{2}\right]+\mu_k\lvert\beta\rvert\omega_k t\cos\left[\frac{(2\omega_k-\omega)t-2\theta}{2}\right]\sinc\left[\frac{(2\omega_k-\omega)t}{2}\right]\nonumber\\
&+(-1)^{k+k'}\mu_{k'}\sqrt{\omega_k\omega_{k'}}\lvert\beta\rvert t\left(\sinc\left[\frac{(\omega_k+\omega_{k'}+\omega)t}{2}\right]\cos\left[\frac{(\omega_k+\omega_{k'}+\omega)t+2\theta}{2}\right]\right.\nonumber\\
&+\sinc\left[\frac{(\omega_k-\omega_{k'}+\omega)t}{2}\right]\cos\left[\frac{(\omega_k-\omega_{k'}+\omega)t+2\theta}{2}\right]+\sinc\left[\frac{(\omega_k-\omega_{k'}-\omega)t}{2}\right]\cos\left[\frac{(\omega_k-\omega_{k'}-\omega)t-2\theta}{2}\right]\nonumber\\
&\left.+\sinc\left[\frac{(\omega_k+\omega_{k'}-\omega)t}{2}\right]\cos\left[\frac{(\omega_k+\omega_{k'}-\omega)t+2\theta}{2}\right]+\frac{8}{\lvert\beta\rvert}\int_0^t dt' \xi(t')\cos(\omega_{k}t') \cos(\omega_{k'}t')
\right)\nonumber\\
&-2(-1)^{k+k'}\sqrt{\omega_k\omega_{k'}}\int_0^t dt'\psi_{k'}(t')\big(4\lvert\beta\rvert\cos(\omega_k t')\cos(\omega t'-\theta)+\xi(t')\cos(\omega_k t')\big)\nonumber\\
&\left.+2\omega_k\int_0^tdt'\left[\Lambda_{pk}(t')(\lvert\beta\rvert\cos(\omega t'-\theta)+\xi(t'))+\Lambda_{k}^-(t')\sin(2\omega_kt')\left(2\lvert\beta\rvert\cos(\omega t'-\theta)+\xi(t')\right)\right]\right\}.
\end{align}

\section{Number of photons: second order correction}\label{N:second:order:correction}
The correction to the number of photons up to the second order, when the initial state is $\hat{\rho}(0)=\prod_{n}\lvert 0_n\rangle\langle 0_n\rvert\otimes\hat{\rho}_{\textrm{m}}^{\textrm{(DST)}}$, can be expressed as a combination of terms, collected in Equation~\eqref{N:second:order:tot}. In more details:
\begin{align}
\Delta N^{(2)}_{\beta,k}(t)=&2\sinc^2\left[\frac{(2\omega_k-\omega)t}{2}\right]+2\sinc^2\left[\frac{(2\omega_k+\omega)t}{2}\right]+4\sinc\left[\frac{(2\omega_k+\omega)t}{2}\right]\sinc\left[\frac{(2\omega_k-\omega)t}{2}\right]\cos(\omega t-2\theta)\nonumber\\
&+\sum_{n\neq k}\frac{\omega_{n}}{\omega_k}\left\{\left(\sinc^2\left[\frac{(\omega_k+\omega_{n}-\omega)t}{2}\right]+\sinc^2\left[\frac{(\omega_k+\omega_{n}+\omega)t}{2}\right]\right)\right.\nonumber\\
&\left.+2\sinc\left[\frac{(\omega_k+\omega_{n}-\omega)t}{2}\right]\sinc\left[\frac{(\omega_k+\omega_{n}+\omega)t}{2}\right]\cos(\omega t-2\theta)\right\},
\label{deltaNbeta}
\end{align}
is the contribution stemming from the coherent state of the quantum harmonic oscillator;
\begin{align}
\Delta N^{(2)}_{\textrm{vac, k}}(t)=&2\sinc^2\left[\frac{(2\omega_k+\omega)t}{2}\right]+\sum_{n\neq k}\frac{\omega_{n}}{\omega_k}\sinc\left[\frac{(\omega_k+\omega_{n}+\omega)t}{2}\right],
\end{align}
is the correction due to the quantum vacuum state of the mechanical oscillation of the wall, and represents the contribution to the number of photons due to the fluctuation of the phonon number;
\begin{align}
\Delta N^{(2)}_{\textrm{sq},k}(t)=&2\sinc^2\left[\frac{(2\omega_k+\omega)t}{2}\right]\sinh r\nonumber+2\sinc^2\left[\frac{(2\omega_k-\omega)t}{2}\right]\sinh r\\
&-4\cosh r\sinc\left[\frac{(2\omega_k+\omega)t}{2}\right]\sinc\left[\frac{(2\omega_k-\omega)t}{2}\right]\cos(\omega t-\phi)\nonumber\\
&+\sum_{n\neq k}\frac{\omega_{n}}{\omega_k}\left\{\left(\sinc^2\left[\frac{(\omega_k+\omega_{n}-\omega)t}{2}\right]+\sinc^2\left[\frac{(\omega_k+\omega_{n}+\omega)t}{2}\right]\right)\sinh r\right.\nonumber\\
&\left.-2\sinc\left[\frac{(\omega_k+\omega_{n}-\omega)t}{2}\right]\sinc\left[\frac{(\omega_k+\omega_{n}+\omega)t}{2}\right]\cosh r\cos(\omega t-\phi)\right\}\nonumber\\
\Delta N^{(2)}_{T,k}(t)=&2\sinc^2\left[\frac{(2\omega_k+\omega)t}{2}\right]+2\sinc^2\left[\frac{(2\omega_k-\omega)t}{2}\right]\nonumber\\
&+\sum_{n\neq k}\frac{\omega_{n}}{\omega_k}\left(\sinc^2\left[\frac{(\omega_k+\omega_{n}-\omega)t}{2}\right]+\sinc^2\left[\frac{(\omega_k+\omega_{n}+\omega)t}{2}\right]\right)\nonumber\\
\Delta N^{(2)}_{\textrm{sq},T,k}(t)=&4\,\sinh r\left[\sinc^2\left[\frac{(2\omega_k+\omega)t}{2}\right]
+\sinc^2\left[\frac{(2\omega_k-\omega)t}{2}\right]\right]\nonumber\\
&-4\,\cosh r\,\sinc\left[\frac{(2\omega_k+\omega)t}{2}\right]\sinc\left[\frac{(2\omega_k-\omega)t}{2}\right]\cos(\omega t-\phi)\nonumber\\
&+\sum_{n\neq k}\frac{\omega_{n}}{\omega_k}\left\{2\,\sinh r\left(\sinc^2\left[\frac{(\omega_k+\omega_{n}-\omega)t}{2}\right]+\sinc^2\left[\frac{(\omega_k+\omega_{n}+\omega)t}{2}\right]\right)\right.\nonumber\\
&\left.-4\,\cosh r\sinc\left[\frac{(\omega_k+\omega_{n}-\omega)t}{2}\right]\sinc\left[\frac{(\omega_k+\omega_{n}+\omega)t}{2}\right]\cos(\omega t-\phi)\right\},
\end{align}
are the corrections due to squeezing alone, initial temperature in the mechanical element alone, and the interaction of the two; and finally
\begin{align}
N_{\textrm{md}}^{(2)}(t)=&2\epsilon^2\omega_k^2 t\left\{\beta\int_0^tdt'\xi(t')e^{-2i\omega_kt'}\left[\sinc\left[\frac{(2\omega_k+\omega)t}{2}\right]e^{i\frac{(2\omega_k+\omega)t}{2}}+\sinc\left[\frac{(2\omega_k-\omega)t}{2}\right]e^{i\frac{(2\omega_k-\omega)t}{2}}\right]\right.\nonumber\\
&+\left.\beta^*\int_0^tdt'\xi(t')e^{2i\omega_kt'}\left[\sinc\left[\frac{(2\omega_k+\omega)t}{2}\right]e^{-i\frac{(2\omega_k+\omega)t}{2}}+\sinc\left[\frac{(2\omega_k-\omega)t}{2}\right]e^{-i\frac{(2\omega_k-\omega)t}{2}}\right]\right\}\nonumber\\
&+\sum_{n\neq k}\frac{\epsilon^2\omega_k\omega_nt}{2}\left\{\beta\int_0^tdt'\xi(t')e^{-i(\omega_k+\omega_n)t'}\left[\sinc\left[\frac{(\omega_k+\omega_n+\omega)t}{2}\right]e^{i\frac{(\omega_k+\omega_n+\omega)t}{2}}\right.\right.\nonumber\\
&\left.+\sinc\left[\frac{(\omega_k+\omega_n-\omega)t}{2}\right]e^{i\frac{(\omega_k+\omega_n-\omega)t}{2}}\right]\nonumber\\
&+\left.\beta^*\int_0^tdt'\xi(t')e^{i(\omega_k+\omega_n)t'}\left[\sinc\left[\frac{(\omega_k+\omega_n+\omega)t}{2}\right]e^{-i\frac{(\omega_k+\omega_n+\omega)t}{2}}+\sinc\left[\frac{(\omega_k+\omega_n-\omega)t}{2}\right]e^{-i\frac{(\omega_k+\omega_n-\omega)t}{2}}\right]\right\}\nonumber\\
&+4\epsilon^2\omega_k^2\int_0^tdt'\xi(t')e^{-2i\omega_kt'}\int_0^tdt'\xi(t')e^{2i\omega_kt'}+\sum_{n\neq k}\epsilon^2\omega_k\omega_n\int_0^tdt'\xi(t')e^{-i(\omega_k+\omega_n)t'}\int_0^tdt'\xi(t')e^{i(\omega_k+\omega_n)t'},
\end{align}
is the number of photons due to the presence of the mechanical drive.

\end{document}